\documentclass[twocolumn]{emulateapj}

\shorttitle{Old Globular Clusters in Magellanic-Type dIrrs.}
\shortauthors{Georgiev, I. Y. et al.}

\slugcomment{AJ accepted, Feb., 2008}

\usepackage{amssymb}
\usepackage{natbib}
\usepackage{epsfig}
\usepackage{color}

\begin{document}

\title{Old Globular Clusters in Magellanic-Type Dwarf Irregular
  Galaxies\altaffilmark{\mbox{$\star$}}} 

\author{Iskren Y. Georgiev\altaffilmark{1,2 \mbox{$\star\star$}}, Paul
  Goudfrooij\altaffilmark{1}, Thomas H. Puzia\altaffilmark{3} and
  Michael Hilker\altaffilmark{4}} 
\altaffiltext{1}{Space Telescope Science Institute, 3700 San Martin Drive, Baltimore, MD 21218, USA}
\altaffiltext{2}{Argelander Institute f\"ur Astronomie der Universit\"at Bonn, Auf dem H\"ugel 71,D-53121 Bonn, Germany}
\altaffiltext{3}{Herzberg Insitute for Astrophysics, 5071 West Saanich Road, Victoria, BC V9E 2E7, Canada}
\altaffiltext{4}{European Southern Observatory, Karl-Schwarzschild-Str.~2, 85748 Garching bei M\"unchen, Germany}

\email{$^{\star\star}$georgiev@stsci.edu, iskren@astro.uni-bonn.de}

\altaffiltext{$\star$}{Based on archival data of the NASA/ESA {\it
    Hubble Space Telescope}, which is operated by AURA, Inc., under
  NASA contract NAS 5--26555.} 

\begin{abstract}
We have performed a search for old globular clusters (GC) using archival $F606W$ and $F814W$ HST/ACS images of 19 Magellanic-type dwarf Irregular (dIrr) galaxies. Those dIrrs reside in nearby (2 - 8\,Mpc) associations of only dwarf galaxies. All dIrrs have absolute magnitudes fainter than or equal to the SMC (M$_V=-16.2$\,mag). We detect in total 50 GC candidates in 13 dIrrs, of which 37 have $(V-I)$ colors consistent with ``blue'' (old, metal-poor) GCs (bGC). The luminosity function (LF) of the bGC candidates in our sample shows a turnover magnitude of $M_V=-7.41\pm0.22$\,mag, consistent with other galaxy types. The width of the LF is $\sigma=1.79\pm0.31$ which is typical for dIrrs, but broader than the typical width in massive galaxies. The half-light radii and ellipticities of the GCs in our sample ($\bar{r_{\rm h}}\simeq3.3$\,pc , $\overline{\epsilon}\simeq0.1$) are similar to those of old GCs in the Magellanic Clouds and to those of ``Old Halo'' (OH) GCs in our Galaxy, but not as extended and spherical as the Galactic ``Young Halo'' (YH) GCs ($\bar{r_{\rm h}}\simeq7.7$\,pc , $\overline{\epsilon}\simeq0.06$). The $\epsilon$ distribution shows a turnover rather than a power law as observed for the Galactic GCs. This might suggest that GCs in dIrrs are kinematically young and not fully relaxed yet. The present-day specific frequencies of GCs ($S_N$) in the galaxies in our sample span a broad range:\ $0.3<S_N<11$. Assuming a dissipationless age fading of the galaxy light, the $S_N$ values would increase by a factor of $\sim2.5$ to 16, comparable with values for early-type dwarfs (dE/dSphs). A bright central GC candidate, similar to nuclear clusters of dEs, is observed in one of our dIrrs: NGC~1959. This nuclear GC has luminosity, color, and structural parameters similar to that of $\omega$\,Cen and M\,54, suggesting that the latter might have their origin in the central regions of similar Galactic building blocks as the dIrrs in this study. A comparison between properties of bGCs and Galactic YH GCs, suspected to have originated from similar dIrrs, is performed. 
\end{abstract}


\section{Introduction}

Hierarchical structure formation models envision the assembly of present-day massive galaxies via numerous minor mergers/accretions of smaller galactic
entities. Present day dwarf galaxies have masses that are similar to those predicted for the proto-galactic fragments \cite[e.g.][]{Purcell07}, which
were later incorporated into more massive galaxies \cite[e.g.][]{Searl&Zinn78}. The prediction of the hierarchical growth of major galaxies through merging of many dwarf-sized fragments at early times is manifested by the steepening of the faint-end slope of the galaxy luminosity function with redshift \cite[e.g.][]{Ryan07, Khochfar07}. Furthermore, dwarf irregular galaxies are found to be one of the most abundant galaxy types in the high-z Universe \cite[e.g.][]{Ellis98, Stiavelli04}, and thus are regarded as building blocks of massive galaxies. As nearby dIrrs are less evolved systems than nearby massive galaxies, they might represent the most ``pristine'' dwarfs that are likely similar to the
fragments incorporated into massive halos and therefore hold important clues as to how galaxies formed.

Old globular clusters (GC) are among the first objects to form in the early Universe. As luminous agglomerations of coeval stars with homogeneous metal abundances they represent unique tools for tracing the main galaxies' star formation episodes. Therefore, their properties, as distinct entities and as a system, reflect the physical conditions at the time of their early formation. This will provide us with constraints on the galaxies' early assembly, since dIrrs may well contribute to the assembly of the rich GC systems (GCSs) of the most massive galaxies.

Numerous observations of the most massive elliptical and spiral galaxies in various environments have shown that they contain exceedingly rich populations of old GCs \cite[e.g.][]{Kundu99, Larsen01, Dirsch03, Harris06a, Tamura06a}. Extensive studies were performed in the last decade to understand how such populous GCSs were assembled \cite[e.g.][]{Ashman&Zepf92, Zepf&Ashman93, Hilker99, Goudfrooij03, Chandar04, Puzia04, Rhode05, Goudfrooij07}. These studies led to two important discoveries: i) the bimodal metallicity/color distribution of the GCs \cite[e.g.][]{Ashman&Zepf92, Neilsen&Tsvetanov99, Gebhardt&Kissler-Patig99, Puzia99, Kundu&Whitmore01} and ii) the presence of young/intermediate age massive star clusters in merging, starburst and irregular galaxies \cite[e.g.][]{Whitmore&Schweizer95, Goudfrooij01, Puzia02, Goudfrooij04}. As a result, three major galaxy/GCS assembly scenarios have been proposed to explain these findings. The first scenario is the {\it hierarchical} build-up of massive galaxies through merging and accretion of pre-galactic dwarf-sized gas fragments \cite[]{Searl&Zinn78} in which the metal-poor GCs form {\it in situ} while the metal-rich GCs originate from a second major star formation event \cite[]{Forbes97} from infalling gaseous fragments, e.g. the mini-mergers at high redshifts \cite[]{Beasley02}. The spiral-spiral dissipative galaxy {\it merger} scenario \cite[]{Schweizer87, Ashman&Zepf92} assumes that the metal-poor GCs were formed early in the progenitor galaxies, while metal-rich GCs formed during the major merger events. Finally, the {\it accretion} scenario incorporates the classical monolithic collapse picture in which the galaxies and their GCs form in the initial starburst \cite[]{Pipino07}and the later accretion of smaller dwarf-sized systems contributed to the assembly of massive galaxies GCSs \cite[]{Cote98, Hilker99, Cote02, Beasley02, Kravtsov&Gnedin05}. All three scenarios are not mutually exclusive and actually bear a number of similarities; they vary mostly in the amount of gas involved. For extensive discussion on the strengths and weaknesses of the various scenarios we refer the reader to the reviews by \cite{Ashman&Zepf98, Kissler-Patig00, vdBergh00, West04, Brodie06}.

Dwarf irregular galaxies play a key role in each GCS/galaxy formation scenario. As dIrrs host mainly old metal-poor, blue GCs (bGCs), the GCSs of massive galaxies, and mainly their bGCs, might be a mixture of clusters formed {\it in situ} and GCs accreted with the dwarfs through hierarchical merging. Thus, it is very important to study dIrr galaxies to obtain a proper understanding of their GCS properties and what might be their contribution to the blue GC population of massive galaxies, which has so far not yet been possible due to the lack of large enough samples.

Analyzing GCs' metallicities and horizontal branch (HB) morphologies, \cite{Zinn93} proposed a sub-classification of the Milky Way (MW) GCS into three populations -- BD (bulge/disk), OH (old halo) and YH (young halo), where BD are all metal-rich ([Fe/H]\,$>-0.8$) clusters, and the metal-poor, OH and YH, are divided by their horizontal branch index, i.e. $\Delta$HB\,$>-0.3$ and $\Delta$HB$<-0.3$, respectively. With the idea that the YHs might be of external origin, first \cite{Zinn93} and later \cite{Mackey&Gilmore04} and \cite{Mackey&vdBergh05} showed that in many respects the properties (colors, luminosities, spatial distributions, structural parameters, HB morphologies etc.) of the MW YH GCs are similar to the properties of GCs in dwarf galaxies (LMC, SMC, Fornax, Sagittarius). Recently, \cite{Lee07} showed that GCs with extended horizontal branches (EHBs) are decoupled in their orbital dynamics and mass from the rest of the GCs and confirmed the \cite{Zinn93} conclusion that the YH clusters are dominated by random motions and share the hot kinematic properties of EHB Galactic GC population supporting their probable external origin. This, and evidences of tidal streams around the MW and M31 \cite[e.g.][]{Ibata01, Martin04, Kalirai06} further supported the view by \cite{Searl&Zinn78} that a fraction of the GCs were formed in isolated low-luminous satellite building blocks, which were later accreted.

The present study focuses on the GC properties (luminosity, color, structural parameters, spatial distribution and specific frequencies) of dIrr galaxies in groups and associations of only dwarf galaxies. The observational data, its reduction, GC selection and photometry are described in Sect.\,\ref{observations}. The results are presented and summarised in Sect.\,\ref{results}, where the properties of GC candidates are discussed and compared with the Galactic YH GCs. Additionally, we address the question whether these dIrrs can be the progenitors of the present day dE/dSph by comparing their GC specific frequencies ($S_N$) with those of early-type dwarfs in clusters \cite[e.g.][]{Miller98,Seth04, Forbes05, Strader06, Miller&Lotz07}.

\section{Observations}\label{observations}

\subsection{HST/ACS data}\label{data}

The current study uses HST archival data of 20 Magellanic-type dwarf irregular (dIrr) galaxies located in nearby ($d=3-8$\,Mpc) loose groups(associations) of dwarf galaxies \cite[]{Tully06}. They were observed in the F606W and F814W filters with HST/ACS during Cycle 13, program GO-10210 (PI:\,B.\,Tully). The split into two non-dithered exposures was designed to reach $\sim2$\,mag beyond the Tip of the Red Giant Branch (TRGB) luminosity at $M_{I}=-4.05, (M_{V}=-2.5)$ \cite[]{DaCosta&Armandroff90, Lee93} in order to derive the TRGB distances to these groups/associations of (dIrr) galaxies. Accurate distances to these galaxies are published in \cite{Tully06}, which provides us with an excellent tool to study the properties of the GCSs of these dwarfs.

\subsection{Initial photometry}\label{ini}

The retrieved archival images were processed with the standard HST/ACS archival pipeline. In order to improve the object detection and initial photometry, the diffuse galaxy light was modeled and subtracted by convolving the images with ring aperture median kernel of 41\,pixel radius. Then, on the residual images we used the {\sc IRAF\footnote{IRAF is distributed by the National Optical Astronomy Observatories, which are operated by the Association of Universities for Research in Astronomy, Inc., under cooperative agreement with the National Science Foundation.} DAOPHOT/DAOFIND} routine to detect objects with $>4$-$\sigma$ above the background. Photometry for all detected sources was performed with the {\sc DAOPHOT/PHOT} routine in apertures of 2,\,3,\,5 and 10 pixel radius. To convert the instrumental magnitudes to the ST magnitude system an aperture correction from 10 pixel aperture to infinity was applied, using the zero points and aperture correction values from the ACS photometric calibration by \cite{Sirianni05}. The magnitudes derived this way are labelled in the following with the subscript `$10\rm tot$'. These magnitudes were only used for the GC selection, while the final GC magnitudes were determined as described in Section\,\ref{gcphot}.

\subsection{Aperture corrections}\label{appcor}

The typical size of a GC, containing half of its light, is about 3\,pc. Therefore, the GCs in the galaxies in our sample are expected to be resolved on ACS images, and typically their light extends beyond 10 pixels, which can introduce systematic errors in the aperture correction. We explore how this error affects our GC selection criteria.
\begin{figure}
\epsfig{file=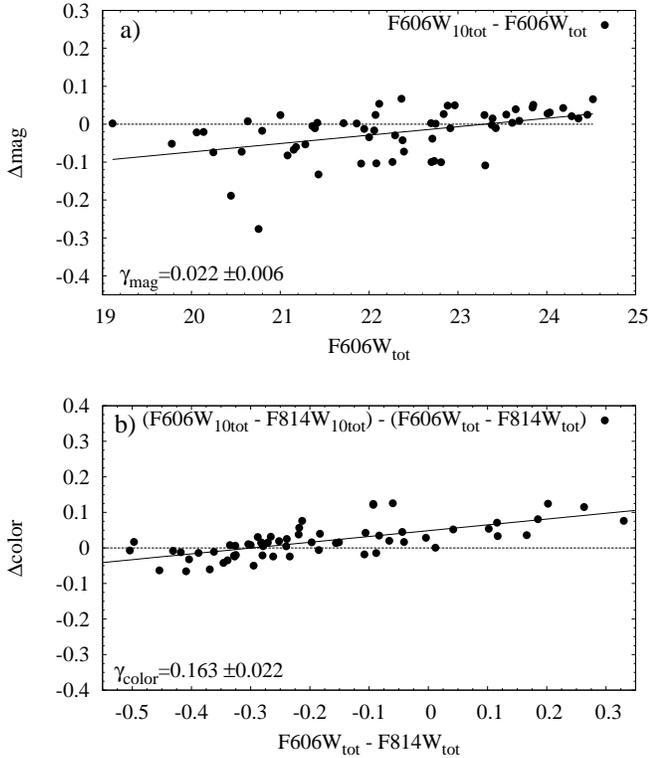,width=.5\textwidth}
\caption{Constant aperture correction is color and magnitude dependent. In panel\,a) are shown the magnitude difference derived from the individual curves of growth for each GC candidate (F606W$_{\rm tot}$ magnitudes) versus the magnitudes simply corrected with a constant value from 10 pixels to infinity (F606W$_{10\rm tot}$). Analogously, in panel\,b) we show the same relation for the objects' colors. The thick line is the least-square linear fit to the data and $\gamma$ is the slope of the fit. It is seen that, even when excluding the faintest objects, which are also the reddest (F606W$_{\rm tot}$\,--\,F814W$_{\rm tot})>0.1$), a simple aperture correction is magnitude and color dependent.\label{ap_diff}}
\end{figure}
In Figure\,\ref{ap_diff} are shown the differences between the corrected (F606W$_{10\rm tot}$; see above) aperture magnitudes (Fig.\,\ref{ap_diff}\,a) and colors (Fig.\,\ref{ap_diff}\,b) and the ``true'' magnitudes (F606W$_{\rm tot}$) derived from curves of growth for each GC (see Sect.\,\ref{gcphot}). The \cite{Sirianni05} aperture corrections are computed for point sources, however, in the case of resolved GCs this correction will be insufficient and introduce an offset in the GCs' magnitudes. Figure\,\ref{ap_diff} shows that this effect appears to be magnitude and color dependent. The dashed lines indicate the zero offset, the thick lines a linear fit to the data and $\gamma$ is the slope of the fit. The observed magnitude dependence can be understood in the sense that the brighter the object the more extended it is. Thus, the correction is underestimated for brighter GCs while it is overestimated for the fainter ones, when a constant aperture correction was applied to all GC candidates. The color dependence of the aperture correction is even more significant (Fig.\,\ref{ap_diff}\,b). Even when the reddest ((F606W$_{10\rm tot}$--F814W$_{10\rm tot})>0.1$) objects are excluded as likely background galaxies, the color dependence does not change significantly. Hence, a constant aperture correction could affect the clusters' colors by up to 0.1\,mag. This result has significant implications for studies dealing with similar quality data.

\subsection{GC Candidate Selection}\label{gcselect}

\begin{figure}
\epsfig{file=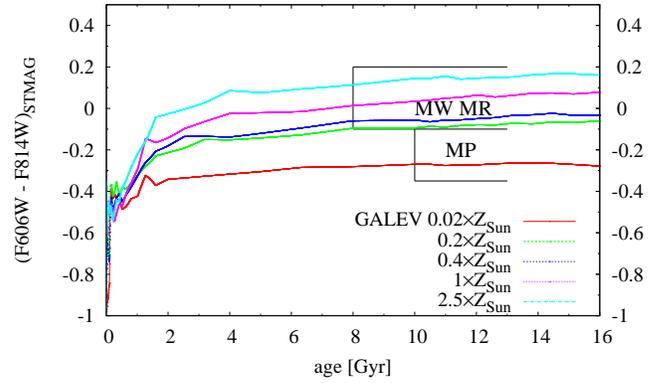,width=.5\textwidth}
\caption{The initial GC selection was based on SSP evolutionary models in the ACS/WFC filter system for different metallicities \protect\cite[]{Anders03}. Namely, objects with colors $-0.4\lesssim$(F606W-F814W)$\lesssim0.2$, i.e. objects older than $\gtrsim4$\,Gyr for the whole range of metallicities. With rectangles are shown the typical color ranges for Galactic metal-poor (MP) and metal-rich (MR) GCs.\label{ssps}}
\end{figure}

Our goal is to study the old GC population in these dIrrs. Thus, our GC candidates selection criteria targets GCs with ages older than $\sim4$\,Gyr for a range of metallicities as provided by the GALEV Simple Stellar Population (SSP) models in the ACS filter system \citep{Anders03}. Therefore, we restrict the color selection limits to $-0.4\lesssim({\rm F606W}-{\rm F814W})_{\rm STMAG}\lesssim0.15$. As seen in Fig.\,\ref{ssps}, the limit at (F606W-F814W$)_{\rm STMAG}\approx-0.4$\,mag (corresponding to Johnson/Cousins $V-I\approx0.7$) is 0.1\,STMAG bluer than the most metal-poor GALEV model. With rectangles in Fig.\,\ref{ssps} are shown the typical color ranges for metal-poor (MP) and metal-rich (MR) Milky Way GCs. Comparison with \cite{BC03} SSP models show that evolutionary tracks with $0.02\times Z_{\odot}$ and $0.005\times Z_{\odot}$ have $V-I$ colors literally indistinguishable for ages older than 4\,Gyr. Therefore, our color limit at (F606W-F814W$)_{\rm STMAG}>-0.4$ selects GC candidates with ages older than 4\,Gyr and $(V-I) > 0.7$\,mag. Considering the weak dependence of $(V-I)$ colors on metallicity, our GC candidates might be as metal-poor as $0.005\times Z_{\odot}$.

Since our GC selection is based on 
%`$10{\rm tot}$' total magnitudes, the adopted color and magnitude ranges account for the
effects mentioned in Sect.\,\ref{appcor} as well. Given the small dispersion in the faint magnitude bins seen in Fig.\,\ref{ap_diff}\,a, the effect of the aperture correction on the initial GC selection is negligible. Therefore, we adopt a faint magnitude selection limit at the TRGB, i.e. $M_{V}=-2.5$ ($M_{I}=-4.05$) \cite[]{DaCosta&Armandroff90,Lee93}, which is $\sim5$\,mag fainter than the typical GC luminosity function turnover magnitude at $M_{V,\,TO}=-7.4$\,mag \cite[e.g.][]{Harris96, Harris01}. This absolute magnitude was converted to apparent magnitudes using the distance modulus for each galaxy as derived by \cite{Tully06}. 

To correct for foreground Galactic extinction we have used the $E(B-V)$ values towards each galaxy from the \cite{Schlegel98} dust maps\footnote{http://nedwww.ipac.caltech.edu/}. We used the Galactic extinction laws by \cite{Cardelli89} (their eqs. 2 and 3) to compute the values for the ACS filters.

\begin{figure*}
\begin{center}
\epsfig{file=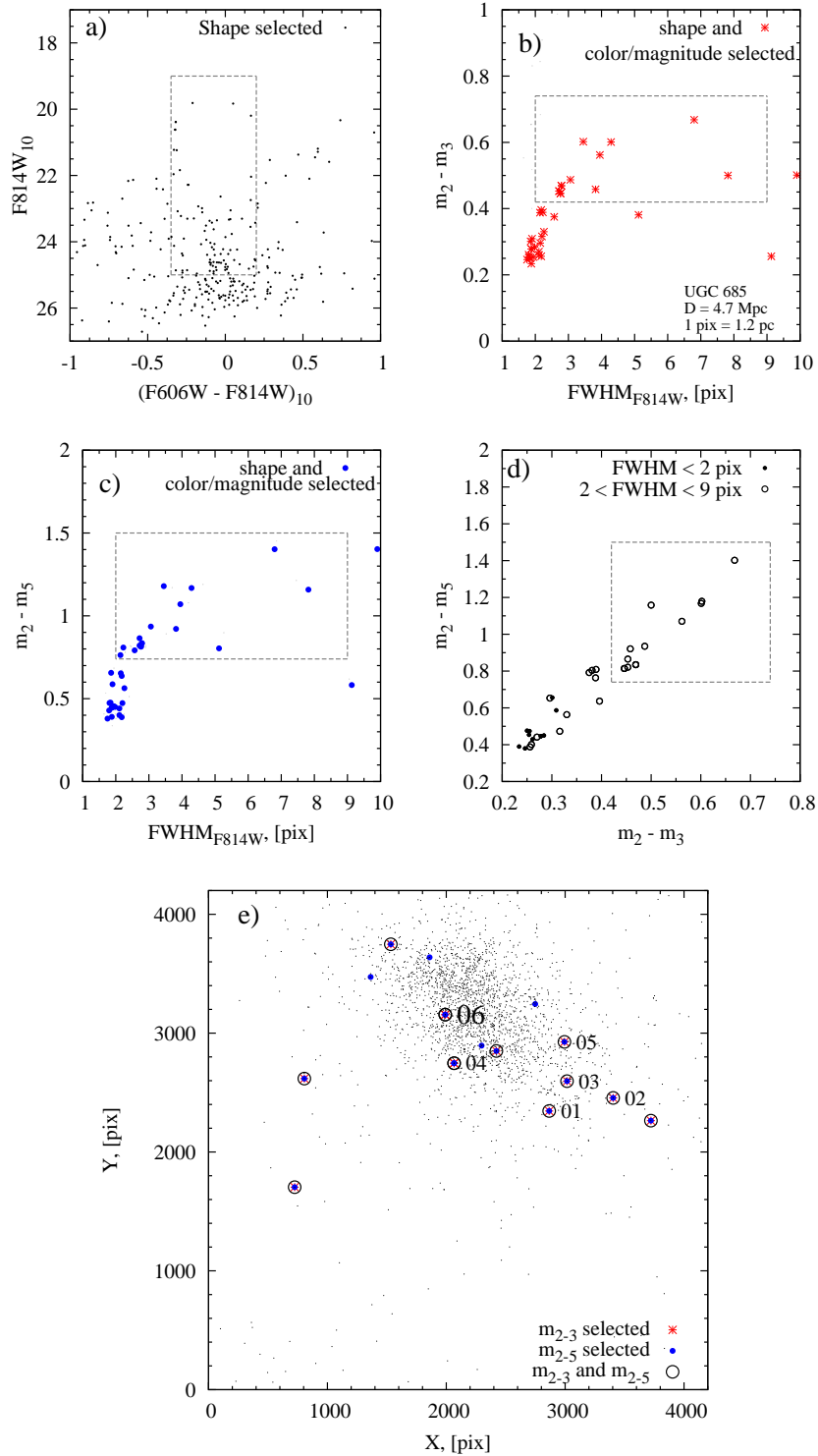, width=.63\textwidth}
\caption{Summary of the initial GC selection process for UGC\,685. In Panel a) are shown the color and magnitude selections. They are indicated by a rectangular region (dashed box) for all objects obeying the shape selection criteria ({\sc imexam} $\epsilon<0.15; {\rm FWHM}_{F606W}\simeq{\rm FWHM}_{F814W}$). In panels b) and c) are shown the objects selected this way which were analyzed against their FWHM and light concentration (both m$_2-{\rm m}_3$ and m$_2-{\rm m}_5$). In panel d) are shown the two concentration parameters against each other. The objects within the dashed box were selected as GC candidates. Panel e): the selected GC candidates (circled) are shown with different symbol types. The candidates which after visual inspection were regarded as GCs are labeled. See Sect.\,\ref{gcselect} for detailed explanation.\label{select}}
\end{center}
\end{figure*}
As mentioned above, due to the deep imaging and high ACS spatial resolution, the GCs in our images show profiles significantly more extended than point
sources (stars). In order to include this knowledge in the GC selection we used {\sc imexam} to obtain initial information on the profiles of all detected objects as first approximation for the GC selection. We considered as GC candidates round objects (FWHM$_{F606W}\simeq$FWHM$_{F814W}$, ellipticity$\leq0.15$) with $2\lesssim$\,FWHM\,$\lesssim9$\,pix (the measured stellar FWHM was $\sim1.7$ to $1.9$\,pix). The {\sc imexam} $\epsilon$ and FWHMs were measured with fixed $r=5$\,pix aperture radius and Moffat index $\beta=2.5$, typical for stellar profiles. Hence, their absolute values cannot be taken as final result but rather as {\it upper} limits for the initial selection criteria. The final ellipticities and half-light radii ($\epsilon,\ r_{\rm h}$) were measured with {\sc ishape} (Sect.\,\ref{determ}). As seen in Fig.\,\ref{ell} and \ref{ell_hist}, the derived $\epsilon$ values for final GC candidates reach 0.3, a typical upper value observed for old GCs in the Magellanic Clouds and our Galaxy (Fig.\,\ref{mwgc_ell},\,\ref{ell_hist}). Therefore, the adopted for initial selection {\sc imexam} $\epsilon<0.15$ within $r=5$\,pix, should not have introduced an artificial bias in our selection, as can be expected for such resolved objects. An additional check, with an increased $\epsilon$ selection limit, proved to include large number of contamination due to blends and nuclei of background galaxies.

The selection process is illustrated in Figure\,\ref{select}, with UGC\,685 as an example. In Fig.\,\ref{select}\,a) we show the color--magnitude diagram (CMD) for all UGC\,685 objects satisfying the aforementioned size selection criteria. The dashed box indicates the color and magnitude selection limits (see above). The observed dispersion in our CMD is simply due to crowding effects leading to the uncertainties in the 10\,pix photometry. Hence, stars affected by their neighbors in crowded regions will show `wrong' colors and increased color dispersion.

As the GCs are expected to show concentrations different to stars and background galaxies, we incorporate this information by calculating the difference between the 2, 3 and 5 pixel aperture photometry. In terms of contamination, the m$_{2}-$m$_3$ index is the more robust star--GC--background
galaxy discriminator than the m$_{2}-$m$_5$ index. This is due to the increase of contaminating light from resolved Galactic stars with increasing aperture size. The GCs occupy a specific locus in the FWHM vs. m$_{2}-$m$_3$ or m$_{2}-$m$_5$ plane, typical for resolved objects. In Fig.\,\ref{select}\,b) and c) we show the concentration m$_{2}-$m$_3$, m$_{2}-$m$_5$ indices versus the measured FWHM$_{F814W}$ for all color, magnitude and shape selected objects ({\sc imexam} $e<0.15$, FWHM$_{F606W}\simeq$FWHM$_{F814W}$). The dividing line around m$_{2}-$m$_3\simeq0.4$ and m$_{2}-$m$_5\simeq0.75$\,mag separates stars from GC candidates. However, the bulges of resolved big galaxies and barely resolved compact (FWHM$<9$\,pix) background galaxies fall in the same region as the GCs \cite[see also][]{Puzia04}. In Fig.\,\ref{select}\,d) the two concentration indices are plotted against each other. The dotted box shows the region occupied by the GC candidates. It can be seen that if we were to select GC candidates from the m$_2-$m$_5$ index alone, we would have introduced more contamination (shown with blue dots) from blended sources within the galaxy. The completeness and contamination issue is discussed in more detail in Sect.\,\ref{comcon}.

In the case of UGC\,685, the automated GC selection procedure returned 13 objects, which were visually inspected. 5 obvious background galaxies and 2
blends were removed leaving 6 GC candidates which are shown with numbered IDs in Fig.\,\ref{select}\,e) (see also Fig.\,\ref{excerpts}). Four of the five objects with $\bar {\rm m}_{2-3}\simeq0.4$ and $\bar {\rm m}_{2-5}\simeq0.7$ in Fig.\,\ref{select}\,e) were visually inspected and turned out to be blended sources in the galaxy body, one was a clear background galaxy. The same was the case in the rest of the studied galaxies, therefore we adopted the lower limit for concentration selection at the quoted values above. The same technique was applied to all our sample galaxies. The final GC candidate sample comprises 60 objects in 13 out of the 19 examined dIrr galaxies.

After the presented selection process, the contamination in our GC sample by unresolved background galaxies cannot be ruled out entirely. Completeness, however, is not an issue due to the depth of the images, with $90\%$ point source limiting magnitude at $M_V=-2.5$\, mag. Therefore, we are confident that we do not introduce any unaccounted artificial bias in our GC selection. A similar selection of GC candidates in nearby LSB dwarf galaxies was recently confirmed with 96\% success rate by radial velocities \cite[]{Puzia&Sharina08}.

\subsection{Final Globular Cluster Candidate Photometry}\label{gcphot}

Accurate photometry of the total light of resolved GCs, including virtually all cluster light, was based on individual curves of growth which account for the different GC sizes due to the varying distances to the host galaxy. However, in images where a significant part of the host galaxy stellar component is resolved a special treatment of the contaminating sources (galaxy stars, complex stellar regions, foreground Galactic stars, background galaxies), found in the photometric apertures, is needed. In order to eliminate the contaminating sources and build curves of growth up to 50 pixel aperture radius, we used the SExtractor \cite[]{Bertin&Arnouts96} package. Measuring the GCs' magnitudes with SExtractor turned out to be an unstable task leading to very uncertain measurements especially for objects in crowded regions. Instead, we used the SExtractor option ``CHECKIMAGE\_TYPE -OBJECTS'' to detect and subtract iteratively in two steps contaminating objects having different profiles. For their detection two different convolution filters were used, provided within SExtractor, i.e. gauss\_5.0\_9x9.conv and gauss\_1.5\_3x3.conv.

\begin{figure*}
\epsfig{file=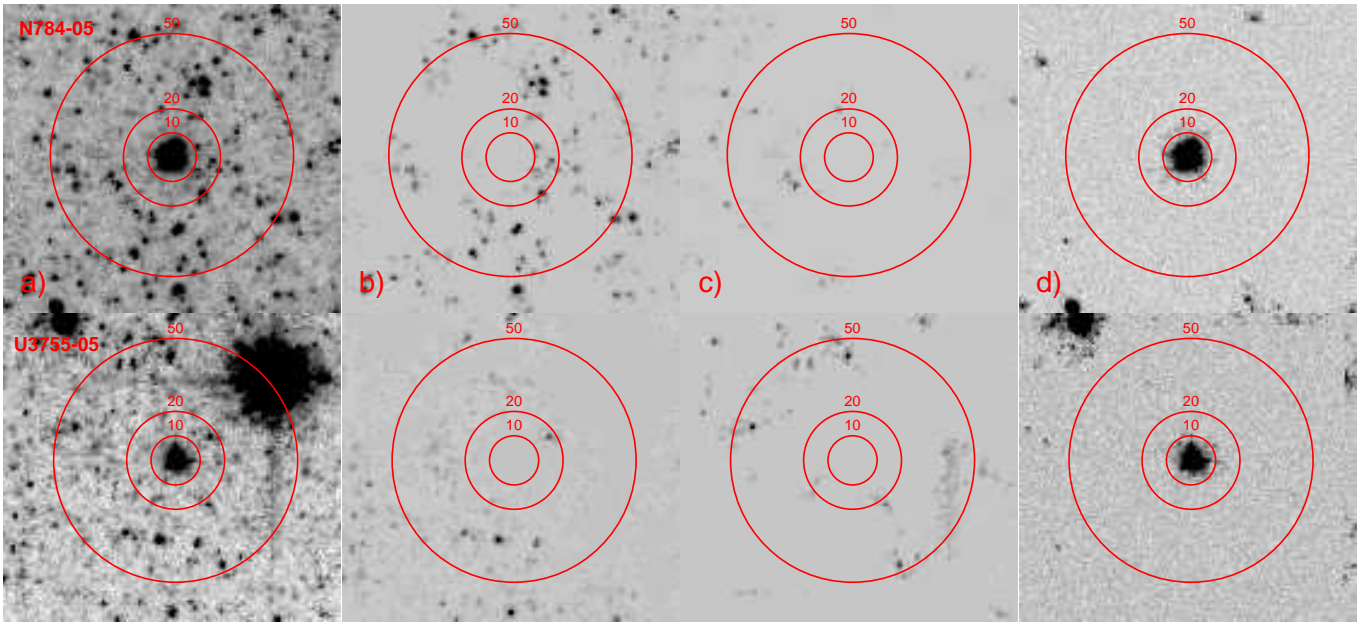,width=1.0\textwidth}
\caption{An illustration of the reduction process of iterative cleaning of the photometric apertures around the GC candidates in two extreme cases with
  strong contamination by resolved galactic stars and a nearby bright foreground star. As discussed in the text, two cleaning iterations (columns b) and c)\,) using SExtractor were involved to remove the underlying galaxy light and contaminating sources within the photometric apertures. The CCD image noise characteristics were restored with {\sc mknoise} (column d). Aperture photometry and curves of growth were build from the images reduced this way.\label{cleaning}}
\end{figure*}
The iterative procedure is illustrated in Figure\,\ref{cleaning} where two cases with strong contamination are shown. In the lower row the brightest star on the right was masked out. The two iterative steps were designed to detect and subtract only sources with the given filter profiles from the images and replace them with the background value as evaluated by a 60-pixel background mesh. This allowed us to adjust the detection thresholds such that (as a first step) the GC with the associated pixels within $0.5\,\sigma$ of the background (thus practically loosing no light from the GC) was subtracted from the original image. The image of this step, containing only contaminating objects, (Fig.\,\ref{cleaning}\,b) was then subtracted from the original (Fig.\,\ref{cleaning}\,a), thus removing only the contaminants. The second iteration used the narrower convolution filter and the object detection was setup such that the GC was detected and subtracted from the result of the first iteration (Fig.\,\ref{cleaning}\,c). Then this image was subtracted from the result of the first iteration, thus removing the last contaminants (see Figure\,\ref{cleaning}\,d). Since SExtractor replaces the detected sources with a constant background value, we used the ACS CCD characteristics from the image header and restored the detector noise in the image with the {\sc IRAF mknoise} routine. This step was mainly performed to restore the proper error budget in the photometry. The circles in Fig.\,\ref{cleaning} mark some of the aperture radii used in our curve of growth photometry.

In order to build individual curves of growth, an aperture photometry from 5 to 50 pixels with a step size of 5 pixels was performed with the {\sc DAOPHOT/PHOT} task on the final images (free of contaminants). Fifth-order polynomial fits to the curves of growth were used to derive the magnitudes for each GC. In order to compare our GC magnitudes with those of other GC studies we iteratively converted the instrumental STMAG magnitudes from ACS/WFC filters to the Johnson/Cousins $V$ and $I$ magnitudes using the synthetic transformation coefficients computed by \cite{Sirianni05}.

\subsection{Globular Cluster Sizes}\label{determ}

With the spatial resolution (0.05\,arcsec/pixel) of the ACS/WFC the GCs in these nearby (2 to 8 Mpc) dIrrs are clearly resolved. This enables one to measure their structural parameters with relatively high precision.

The GC half-light radii (r$_h$) were estimated using the {\sc ishape} algorithm \cite[]{Larsen99}, which models the source as an analytic function convolved with a (model) point-spread function (PSF). For the PSF model we used the {\sc TinyTim} software package\footnote{http://www.stsci.edu/software/tinytim/tinytim.html}. It properly takes into account the field-dependent WFC aberration, filter passband effects, charge diffusion variations and varying pixel area due to the significant field distortion in the ACS field of view \cite[]{Krist&Hook04}. We have used the GCs coordinates to create ten times sub-sampled model PSFs at the position of each GC. When sub-sampling is enabled, {\sc TinyTim} doesn't convolve the PSF with the charge diffusion kernel (CDK), but provides its values in the PSF image header. As it is position and wavelength dependent we generated an individual CDK for each GC, which was later used with {\sc ishape} in the profile fitting process.

Assuming a \cite{King62} model for the intrinsic radial luminosity profile of the clusters, the clusters $F606W$ images were modeled with all concentration parameters $C=r_{t}/r_{c}=5, 15, 30, 100$ provided by {\sc ishape}. An 8\,pixel fitting radius was used. The FWHM of the best $\chi^2$ model was adopted for the objects. To test the stability of the measured FWHMs we varied the fitting radius in the range 4 to 10\,pix corresponding from $\sim6$ to 15\,pc. No significant trend and practically identical FWHMs were returned by {\sc ishape}. In order to convert from FWHM to $r_{\rm h}$ we used the conversion factors tabulated for KING models with different $C$ by \cite{Larsen06}. However, this relation is valid for circularly symmetrical profiles while the output of {\sc ishape} is the FWHM along the major axis ($w_{y}$). In order not to overestimate the $r_{\rm h}$ by simply assuming its measured $r_{{\rm h}, w_{y}}$ value, a correction considering the ratio of the FWHMs along the minor and major axes ($w_{x}/w_{y}$) should be applied. Therefore we adopted for the ``true'' effective radius the geometric mean of the FWHMs along the two axes, i.e.
\begin{equation}
r_{\rm h}=r_{{\rm h}, w_{y}}\sqrt[]{w_{x}/w_{y}}
\end{equation}
where $w_{x}/w_{y}$ is the {\sc ishape} output of the minor/major axes FWHMs ratio. The  comparison between $r_{\rm h}$ values derived this way and the numerically derived correction \cite[eq.\,11 in][]{Larsen06} resulted in absolutely identical values. The analysis and interpretation of the GCs sizes derived as described above is presented in Sect.\,\ref{sizes}.

\subsection{Completeness and Foreground/Background Contamination}\label{comcon}

Given the depth of the imaging data, incompleteness is not an issue since our lower magnitude for GC selection is $\sim2$\,mag above the designed photometric limit of the images for each galaxy (see Sects.\,\ref{data} and \ref{ini}). Further, all dIrr galaxies in our sample have very small radial surface brightness gradients which do not bias the completeness towards the galaxy center and therefore no radial completeness tests are required.

Although the supreme ACS resolution and deep exposures allowed us to reject foreground stars and background galaxies at high confidence, compact background galaxies at intermediate redshifts may share the same color, magnitude and size properties as the GCs \cite[e.g.][]{Puzia04}. In order to assess this type of contamination we used the Hubble Ultra Deep Field (HUDF) images. An ideal way to estimate the contamination from the HUDF is to run the same image reduction, selection and photometry procedures. However, the lack of F814W imaging in HUDF prevents us from performing this task. Therefore, we selected objects having the same characteristics as the GCCs in our F606W images, from the HUDF catalog\footnote{http://heasarc.gsfc.nasa.gov/W3Browse/all/hubbleudf.html}, i.e. $19<V_{\rm F606W}<26$\,mag, ellipticity smaller than 0.15 and $2<$\,FWHM\,$<9$\,pix. Note that $V_{\rm F606W}\sim26$\,mag is equal to the magnitude limit we used for the most distant galaxies in our sample (i.e. UGC\,3974, UGC\,3755 at $\sim8$Mpc), therefore the above set of parameters for selecting objects from the HUDF will be exposure independent. Thus, the objects in the HUDF and our fields should be similarly resolved up to the magnitude limit used for selection.

\begin{figure}
\epsfig{file=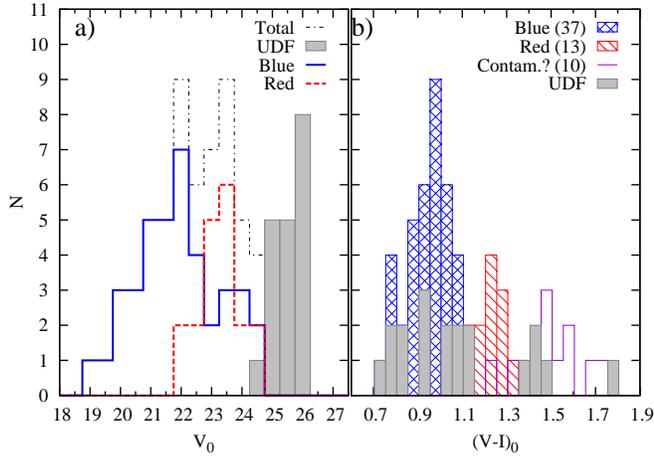,width=.5\textwidth}
\caption{Apparent V-band luminosity (panel a) and color (panel b) distribution of the expected Hubble UDF background contamination (filled histograms). The dashed-dotted line, in panel a) shows the luminosity distribution of all GC candidates in our sample while with thick (blue) and dashed (red) histograms show the candidates divided by their color into blue and red GCs, respectively (Sect.\,\ref{magcol}).\label{contam}}
\end{figure}

After selecting HUDF objects this way, we converted their magnitudes to the Johnson/Cousins system using the \cite{Sirianni05} transformation coefficients and show their magnitude and color distributions in Figure\,\ref{contam}.
The dashed-dotted line, in Fig.\,\ref{contam}\,a), shows the apparent magnitude distribution of all GC candidates in our sample while the solid (blue) and dashed (red) histograms show the candidates divided by their color into blue and red GCCs, respectively (see Sect.\,\ref{magcol}). We overplot the HUDF objects with a shaded histogram. While covering the same $(V-I)$ color distribution as the GC candidates (Fig.\,\ref{contam}\,b), the expected HUDF contamination from background galaxies starts to be a significant factor at faint apparent magnitudes, $V_{0}\gtrsim24$\,mag, as seen in Fig.\,\ref{contam}\,a). Hence, we might expect contamination only in the last one or two magnitude bins. Given that in Fig.\,\ref{contam} we combine the GC candidates from 15 individual ACS fields we need to multiply the number of HUDF contaminating objects by 15. Considering the apparent magnitude distribution of the blue and red GC candidates and the properly scaled HUDF contaminants, our upper estimates are 6 and 9 contaminants, respectively.

\begin{figure}
\epsfig{file=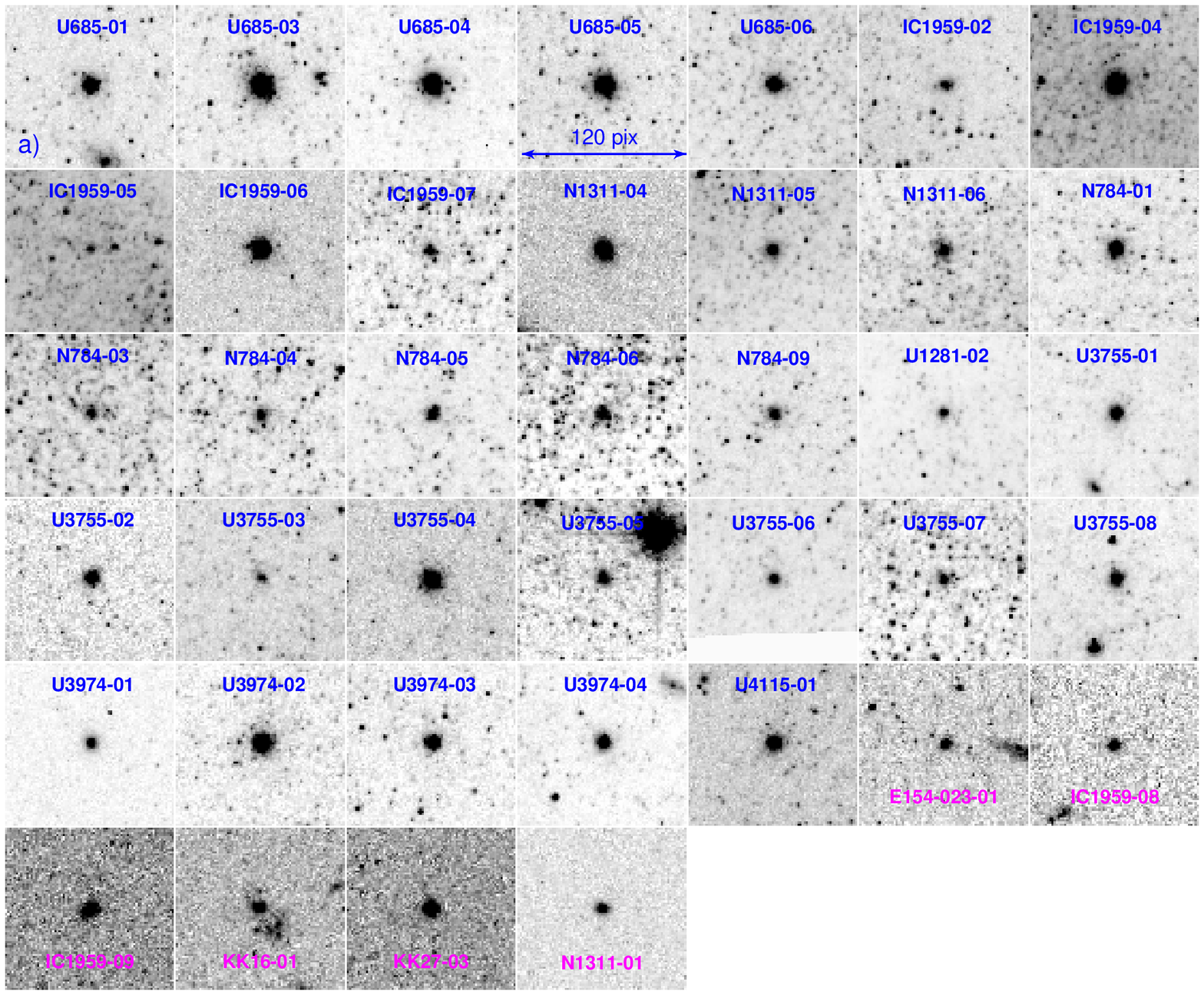,width=0.5\textwidth}
\epsfig{file=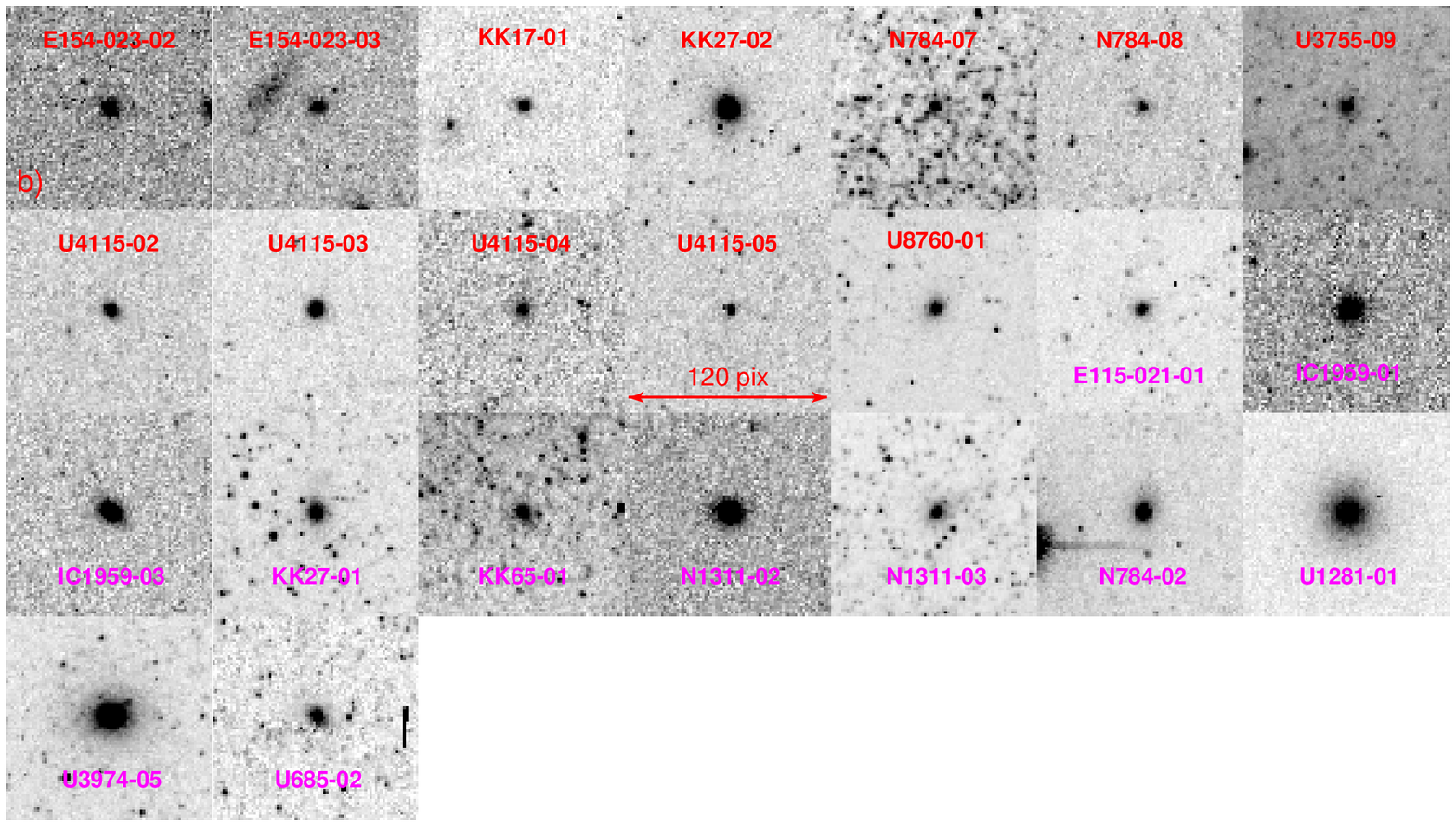,width=0.5\textwidth}
\caption{$120\times120$\,pix F814W excerpts for all GC candidates. Labels on the top of the images indicate the name of the host galaxy and the sequence number of the GC candidate. Panel\,a) shows the GCs with $0.7<(V-I)_{0}<1.15$\,mag. The last eight GCs, having their labels in the bottom of the image, are the faintest and reddest candidates among the blue population (compare with Fig.\,\ref{cmd} and Fig.\,\ref{hist}), thus are likely background contaminants. Panel\,b) shows the red ($(V-I)_{0}\gtrsim1.15$\,mag) group of objects. Although excluded from the analysis, just for comparison we show U\,1281-01 and U\,3974-05, the brightest objects in this group, having extended halos perhaps suggesting that they are background galaxies. This demonstrates that even with the supreme ACS resolution only spectral analysis can resolve all doubts regarding their nature, especially for objects like U\,3974-05.\label{excerpts}}
\end{figure}
In Figure\,\ref{excerpts} we show $120\times120$\,pix F814W image excerpts for all GC candidates. With blue and red labels in Fig.\,\ref{excerpts}\,a),\,b) are indicated the GCs bluer or redder than $(V-I)_{0}=1.1$\,mag, respectively. The objects with labels at the bottom (in magenta) in Fig.\,\ref{excerpts}\,a) and b) are the reddest GCs with $(V-I)_{0}>1.4$\,mag (cf. Fig.\,\ref{cmd} and \ref{hist}) which are very likely background contaminants. Fig.\,\ref{excerpts} shows that the probable contaminants are indistinguishable in appearance from the blue GC candidates in the ACS images.

Red GCs are unlikely to be observed in metal-poor dIrr galaxies because such red metal-rich GCs are believed to be forming mainly in massive galaxies. This question will be further addressed in Sect.\,\ref{results}. Therefore, in the following sections our analysis will be mainly focused on the blue GCCs.

\section{Analysis}\label{results}

\subsection{Magnitudes and Colors}\label{magcol}

\begin{figure}
\epsfig{file=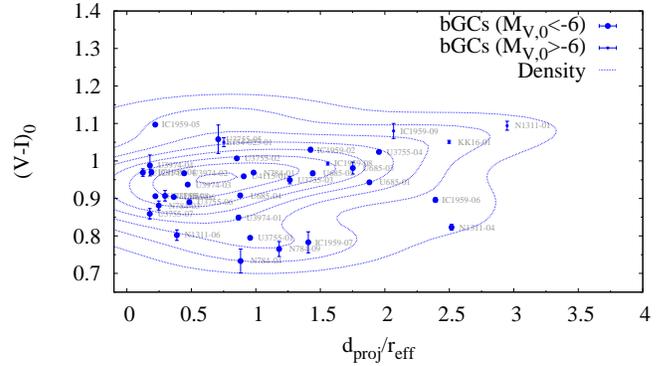,width=.5\textwidth}
\caption{Color vs. projected distance for the blue GCs. The iso-lines of the two dimensional density estimate shows that there is no hint for GCs colors becoming redder with decreasing projected distance, as it might be expected due to internal reddening within the host galaxy.\label{color-dist}}
\end{figure}

Given the low luminosities ($M_V\gtrsim-17$\,mag, see Table\,\ref{main}) of the dwarf galaxies in our sample and the expected low GC specific frequency for dIrr galaxies in group environments \cite[e.g. the Local Group,][]{Harris91} one would expect to find one old GC in each of these galaxies. However, our analysis, shows that several of the faint dwarf galaxies actually harbor more than one GC candidate. As expected, most of the candidates show blue $(V-I)$ colors consistent with old metal-poor GCs.
\begin{figure}
\epsfig{file=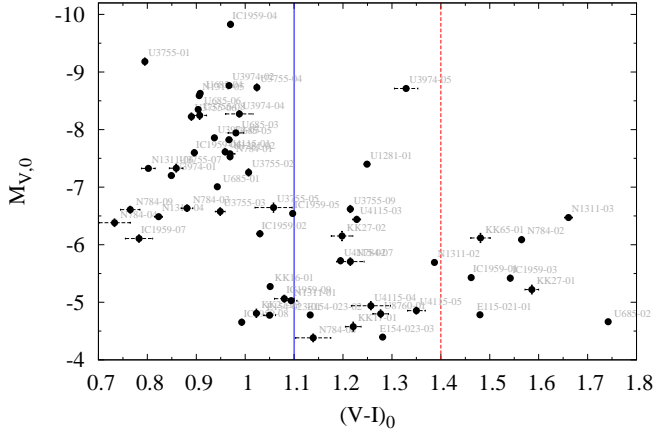,width=.5\textwidth}
\caption{Color-magnitude diagram for all objects that passed the GC selection criteria (Sect.\,\ref{ini}). The colors of the bulk of the GC candidates are consistent with old metal-poor GCs. The labels indicate the host galaxy and the number of the GC candidates in that galaxy. The GC candidates colors and magnitudes were dereddened for foreground Galactic extinction using the \protect\cite{Schlegel98} dust maps. With thick solid and dashed vertical lines are indicated the colors separating blue and red GCs.\label{cmd}} 
\end{figure}
Figure\,\ref{cmd} shows the combined color--magnitude diagram for all GC candidates selected among the 19 dIrrs in our sample (Sect.\,\ref{ini}), that fulfill the color, magnitude, size criteria and passed the additional visual inspection which excluded background galaxies or blends from the final list. Other studies \cite[e.g.][]{Seth04, Sharina05, Georgiev06} of old GCs in dwarf irregular galaxies outside the Local Group showed that the GCs are typically constrained within the color regime $0.8<(V-I)_{0}<1.1$, which corresponds to $>5$\,Gyr old, metal-poor ($Z<0.02\times Z_{\odot}$) GCs according to the SSP models \cite[e.g.][]{BC03}. Compared with the SSP models, the two very blue ($(V-I)_{0}<0.7$) GCCs (NGC\,784-05,06) are likely intermediate-age star clusters with an upper age limit of $\lesssim2$\,Gyr. These two were therefore excluded from our analysis since we are interested only in the old GC population. The population of red objects ($(V-I)_{0}>1.1$) in our CMD, which passed our initial selection criteria, most likely is explained by unresolved, redshifted compact galaxies. Further, such red, possibly metal-rich GCs, are not expected in dIrrs although few of them were observed by \cite{Olsen04} in Sculptor group late-type dwarfs. It is possible that some of the GCs might be reddened, in which case a dependence of their colors with the galactocentric distance would be expected. It is seen from Figure\,\ref{color-dist} that such a dependence is not present.
In fact, we see some evidence for a redder color with increasing projected distance. Given the low number statistics, this trend could be due to unresolved background contaminants in the galaxies' outer regions.

However, follow-up spectroscopy is needed in order to constrain their nature. The two bright and red objects ($M_{V}>-7$\,mag, $1.2<(V-I)_{0}<1.4$) in Fig.\,\ref{cmd} (UGC\,1281-01, UGC\,3974-05) show slightly different (more complex) profiles compared to the rest of the GC candidates and therefore are most probably background contaminants. We therefore also exclude these from subsequent analysis. Objects redder than $(V-I)_{0}>1.5$, although compact and morphologically similar to the bluer GC candidates, are deemed background contaminants. The structural parameters of these objects are discussed in Sect.\,\ref{sizes}.

A detailed examination of Fig.\,\ref{cmd} reveals that there seems to be a lack of faint blue GCs in our CMD ($(V-I)_{0}<1;M_{V}\lesssim-6$\,mag). We have extensively tested the reliability of this observation, by loosening our size/shape selection criteria (the limiting {\sc imexam} ellipticity and FWHM$_{F606W}\simeq$\,FWHM$_{F814W}$ cutoffs). This resulted in an increased number of contaminating blended stars in crowded regions. Therefore this lack of faint bGCs is not due to incompleteness or biased GC selection introduced by our size discrimination criteria (see Sect.\,\ref{ini}). However, we do find GC candidates redder $(V-I)_{0}\sim1$ and fainter than $M_{V}\sim-6$\,mag, which are likely background contaminants. This issue is further discussed in Sect.\,\ref{lumcol-discuss}.

\begin{figure*}
\epsfig{file=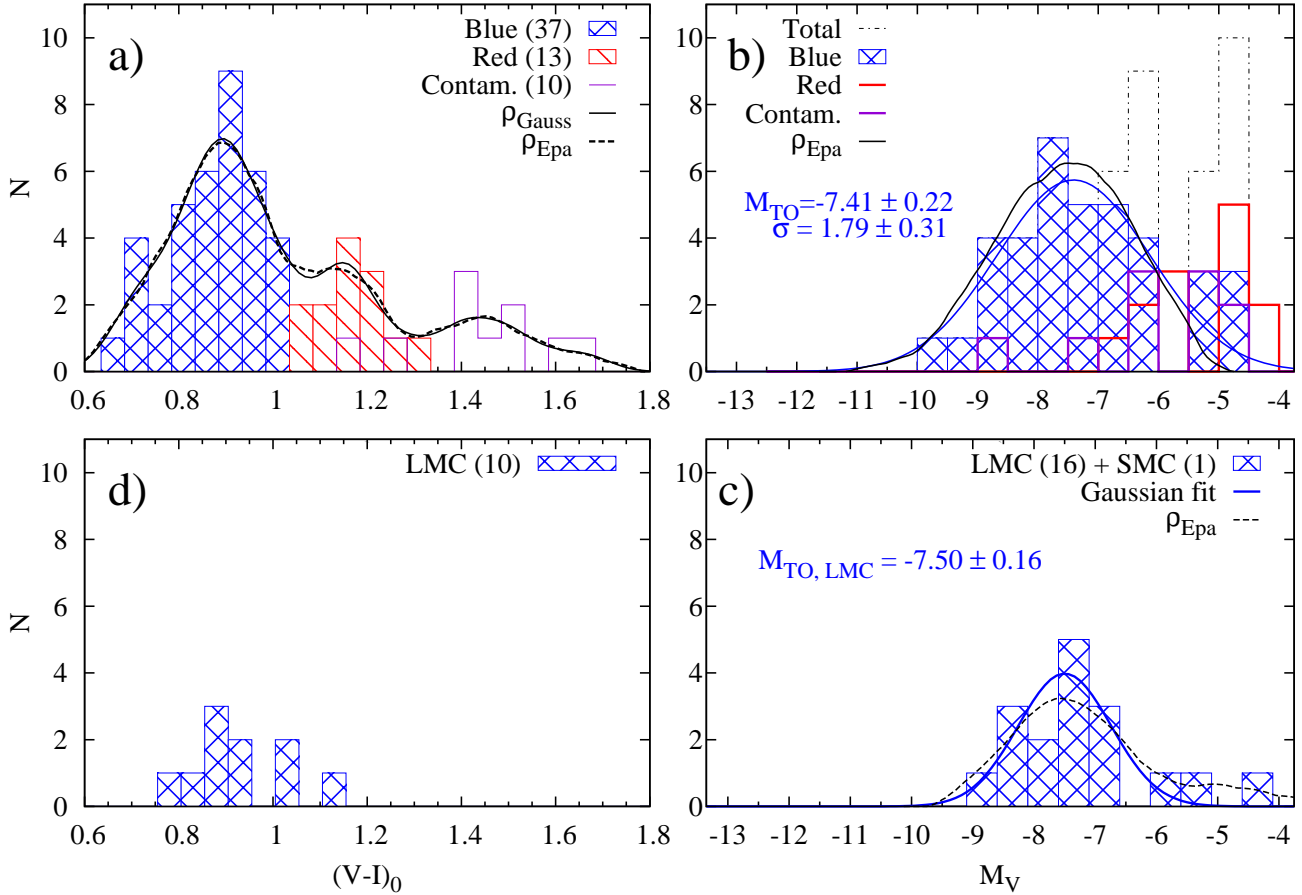,width=1.\textwidth}
\caption{Color and luminosity functions (panels\,a and b) of all GCCs detected in 15 dIrr galaxies. The blue histogram marks the likely GCs, while the likely background contaminants are plotted as red histogram. With thick and dashed lines are shown the non-parametric density estimates using Gaussian and Epanecknikov kernels. Panel\,b) shows the increasing number of objects in the red group with decreasing magnitude, which is mainly due to background contamination (see Sect.\,\ref{comcon} and Fig.\,\ref{contam}). To compare the LF of the GCs in our sample with that of the Magellanic-type dIrrs, panels c) and d) show the LF of old GCs in the LMC, SMC and Fornax dwarf galaxy \protect\cite[data from][]{McLaughlin&vdMarel05}.\label{hist}}
\end{figure*}
In Figure\,\ref{hist} we present the combined luminosity and color distributions of all GC candidates, without the excluded objects as discussed above. In order to probe the underlying distributions we run a non-parametric probability density estimator using Gaussian and Epanechnikov (inverted parabola) kernel, shown with solid and dashed lines in Fig.\,\ref{hist}\,a). Although the Gaussian kernel shows better the dip at $(V-I)_{0}\sim1.1$ between the red and blue objects, both kernels give similar color number density probability distributions, with maxima and minima at the same colors. The third peak represents the population of very red objects, i.e. contaminating background galaxies. The highest probability density peak values are at $(V-I)_{0}\sim0.96$ and $(V-I)_{0}\simeq1.19$ for the blue (bGC) and red GCs, respectively. The peak $(V-I)$ color of the bGCs is similar to the one found in other dIrr galaxies \cite[e.g.][]{Seth04, Sharina05, Georgiev06}. Although our GC sample might have reddening uncertainties, it is anticipated that it is not a significant factor for these extremely low-mass galaxies. Based on $H_{\alpha}$ fluxes \cite{James05} showed that dIrrs posses low $A_{H_{\alpha}}\simeq0.5$, which translates to $E_{B-V}=0.06$ or $E_{V-I}=0.07$\,mag. Note, however, that this only refers to the star forming regions.

The luminosity distribution of the bGCs in our sample peaks at $M_{TO,V}=-7.43$ (rather typical for massive galaxies) and has a broader dispersion than the observed LF for the LMC, SMC and Fornax dwarf galaxies \cite[data from][]{McLaughlin&vdMarel05}, shown in Fig.\,\ref{hist}\,c). The broader luminosity distributions might be partly attributed due to the distance uncertainties to these dwarfs of $\sigma(m-M)\simeq0.05$ \cite[]{Tully06}. A Kolmogorov--Smirnov test gives a 49\% probability that the $M_V$ distribution of our bGCs are drawn from those of the LMC. In comparison with other low-mass dIrr galaxies \cite[$M_{TO,V}\sim-6$;][]{Sharina05, vdBergh06} the bGCs in our sample show a brighter $M_{TO}$.

It is interesting to note that the LMC data from \cite{McLaughlin&vdMarel05}, which is the re-calibrated and extended version of the LMC data from \cite{Mackey&Gilmore03a, Mackey&Gilmore03b}, gives a brighter peak magnitude at $M_{TO,V}=-7.50\pm0.16$\,mag (highest probability density peak at $M_V = -7.52$\,mag) versus $M_{TO,V}=-7.31$\,mag, derived from fitting the formerly determined $M_{V}$ values of old LMC GCs \cite[Table\,2 in][]{vdBergh&Mackey04}. Given the uncertainties, this suggests that the GCLF of old GCs for the dIrrs in our sample, and LMC (considering the uncertainties), both peak at similar magnitudes.

\begin{figure}
\epsfig{file=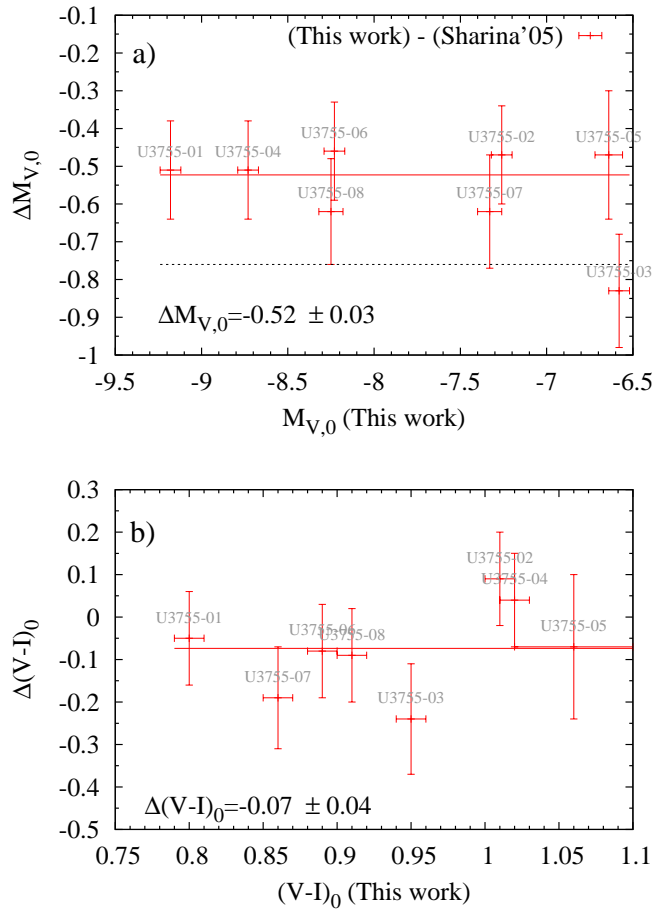, width=0.5\textwidth}
\caption{Differences in absolute magnitudes and colors for our GCs in common with \protect\cite{Sharina05}. The least-square fit to the data (solid line) shows that our magnitudes are 0.5\,mag brighter than the earlier study, while the expected difference due to distance modulus is --0.76\,mag (dotted line). The colors (within the errors) are basically identical between the two studies.\label{diff}}
\end{figure}

Much fainter $M_{TO}$ values were found for other dIrrs by \cite{Sharina05} using WFPC2 data. However, some of the galaxies included in their study (eg. KK\,16, UGC\,8651, UGC\,4115, UGC\,3974, UGC\,3755) were re-observed with much deeper ACS observations by \cite{Tully06}. For these dIrr the distance moduli were improved and some of them turned out to be more distant than thought before. Therefore, underestimating the distance to some galaxies of the \cite{Sharina05} sample led to an underestimate of the GCs' absolute magnitudes, and hence the $M_{TO}$ magnitude. The larger distances also have an effect on the GC half-light radius estimates (see Sect.\,\ref{sizes}). For the GCs in common with the \cite{Sharina05} study in UGC\,3755, the differences in absolute magnitude between both studies are mainly due to the improved photometric technique we used for deriving the clusters apparent magnitudes and the new distance moduli. Figure\,\ref{diff} shows the differences between both studies.
However, the faintness and broadness of the \cite{Sharina05} GC LFs is not expected to be influenced by contamination from background galaxies since their estimated contamination level is less than $\sim5\%$ for objects with $(V-I)_0<1.2,\ -5.3<M_{V_{0}}<-6.3$\,mag \cite[]{Puzia&Sharina08}.

\subsection{Comparison with the Milky Way GCs}\label{MWcomparison}

\begin{figure}
\epsfig{file=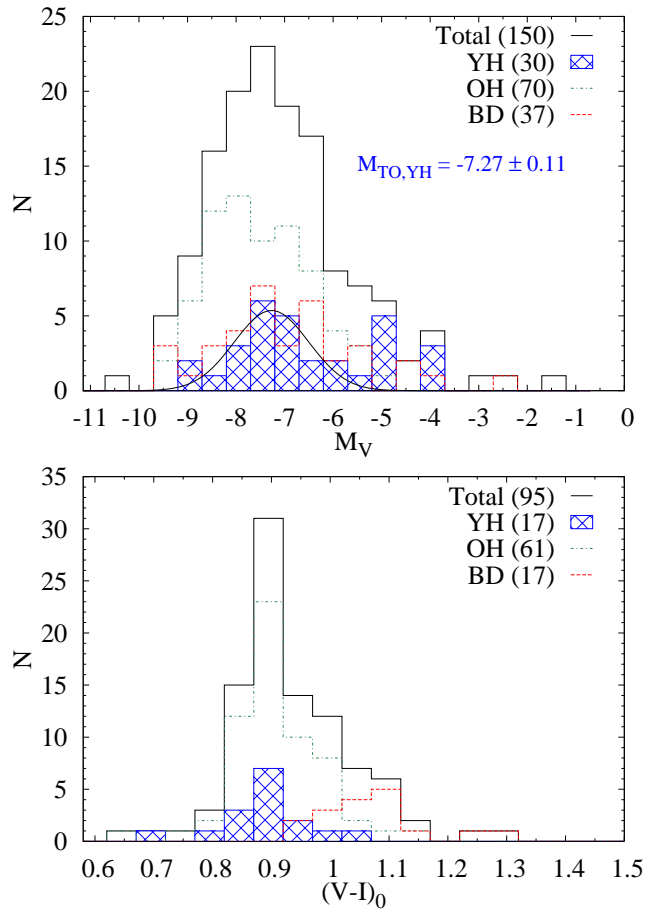,width=0.5\textwidth}
\caption{Color and magnitude distributions of Galactic GCs \protect\cite[data from][revised 2003]{Harris96}. Based on their metallicity and horizontal branch morphology, Bulge/Disk (BD), Old Halo (OH) and ``Young'' Halo (YH) GCs are shown in red, green and blue, respectively. Excluding the last two bins, a Gaussian fit to the YH GC histogram gives a $M_{TO, YH}$ value that is fainter than those of our dIrrs' GCs and the LMC GCs (see Fig.\,\ref{hist}).\label{MWGCcolors}}
\end{figure}
In Figure\,\ref{MWGCcolors} we show the color and magnitude distributions of the different GC subpopulations in our Galaxy. The Milky Way ``Young Halo'' (YH) GCs, which were suggested to have been accreted from dwarf galaxies \cite[e.g.][]{Zinn93, Mackey&Gilmore04, Mackey&vdBergh05,Lee07} show a turnover magnitude ($M_{TO}=-7.27\pm0.11$\,mag) which, although fainter, is consistent within the uncertainties with that of our bGCs ($M_{TO}=-7.41\pm0.22$\,mag). In comparison with the Magellanic Clouds (see also Sect.\,\ref{magcol}), the magnitude distribution of the LMC old GCs peaks at surprisingly bright magnitude $M_{TO}=-7.78\pm0.12$. When including the old GCs of the SMC and the Fornax dSph the combined peak magnitude becomes slightly fainter ($M_{TO}=-7.67\pm0.1$\,mag). K-S tests (cf. Table\,\ref{k-s-results}) assign 54\% and 50\% similarity for the $V-I$ and $M_V$ distributions, respectively, between our bGCs and the Galactic YH clusters. It should be noted, that only 17 YH GCs have $V-I$ colors in the \cite{Harris96} catalog. Comparing with the same distributions for the OH GCs gives probabilities of 2\% and 19\%. The luminosity function of the old LMC clusters is similar to the YH and OH clusters with K-S probability of 43\% and 78\%, respectively. In Table 4 are also shown the results from K-S tests between the $V-I$ and $M_V$ distributions for ten of the old GCs in the LMC with V-I colors \cite[]{McLaughlin&vdMarel05} and bGC in our dIrrs sample.

\subsection{Structural parameters}\label{sizes}

\subsubsection{Half-light radii}\label{rh}

The half-light radius $r_{\rm h}$, a measure of the GC size and stable over many $(>10)$ relaxations times \cite[e.g.][]{Spitzer&Thuam72, Aarseth&Heggie98}, is the most robust and easy to measure structural parameter and therefore is a good indicator for the initial conditions of the GC formation. Hence, a comparison between $r_{\rm h}$ of the blue GCs in our dIrr sample and $r_{\rm h}$ of the YH GCs in our Galaxy and in the LMC can reveal an important evolutionary link among them as well as information of the initial conditions of GC formation.

\begin{figure}
\epsfig{file=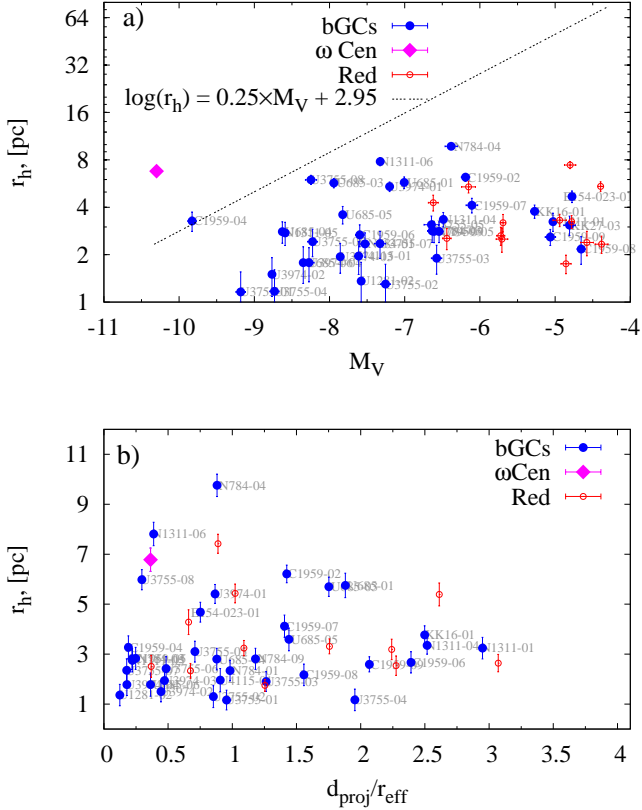,width=0.5\textwidth}
\caption{a) Distribution of the GC candidate sizes ($r_{\rm h}$) with their absolute magnitude. b) $r_{\rm h}$ as function of the normalized projected distance ($d_{\rm proj}/r_{\rm eff}$) from the host galaxy center (b). The GC candidates occupy the $M_V$ vs. $r_{\rm h}$ plane below the relation defined by \protect\cite{Mackey&vdBergh05} (thick line in panel a), which is typical for GCs in our Galaxy and other galaxies. IC\,1959-04, the nuclear cluster of IC\,1959, shares the same parameter space as $\omega$\,Cen.
\label{rhMv}}
\end{figure}
The $r_{\rm h}$ versus $M_V$ relation in Fig.\,\ref{rhMv}\,a) shows that most of our GC candidates fall below the upper envelope of the GC distribution in our Galaxy \cite[solid line in Fig.\,\ref{rhMv}; see ][]{Mackey&vdBergh05}. Interestingly, the brightest cluster, IC1959-04, is at the very center of its host galaxy and occupies the same $r_{\rm h}$ versus $M_V$ region and has similar color as $\omega$\,Cen, which supports a scenario for $\omega$\,Cen's origin as the nucleus of a former dwarf galaxy \cite[e.g.][]{Hilker&Richtler00, Hughes&Wallerstein00}. However, this is the only nuclear cluster we find in the studied galaxies. It is known that $r_{\rm h}$ increases with galactocentric distance for MW, M\,31 and Cen\,A GCs \cite[e.g.][]{Barmby07}, which indicates the influence of the galactic potential. In Fig.\,\ref{rhMv}\,b) we combined the GCs $r_{\rm h}$s from all galaxies in our sample.
Note that in Fig.\,\ref{rhMv}\,b we define $r_{\rm eff}=\sqrt{a \times b}$, where $a$ and $b$ are the galaxy semi-major and semi-minor axes of the galaxy light profile. A dependence of $r_{\rm h}$ on the projected distance from the galaxy center can not be confirmed with significance given the large scatter and the low number statistics.

\begin{figure}
\epsfig{file=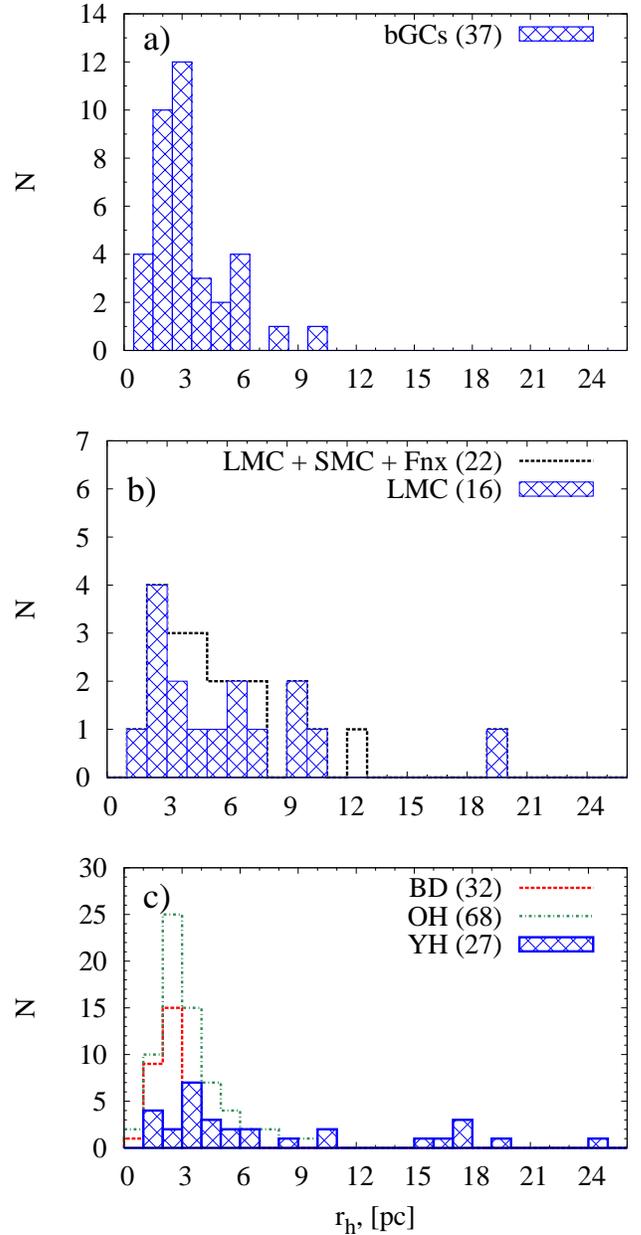,width=0.5\textwidth}
\caption{Comparison of $r_{\rm h}$ distributions for (a) GCs in our dIrrs with that of (b) LMC (+SMC+Fornax) GCs and (c) MW GCs \protect\cite[data from][]{Mackey&vdBergh05, McLaughlin&vdMarel05}.\label{rhHist}}
\end{figure}
In Figure\,\ref{rhHist} we show the size distribution of the GCs in our sample compared with that of the old GCs in the Magellanic Clouds and GCs in our Galaxy. Our bGCs are more compact on average ($\bar r_{\rm h}\sim$3.3\,pc, cf. Fig.\,\ref{rhHist}) than the LMC GCs ($\bar r_{\rm h}\sim6$\,pc) and the YH GCs ($\bar r_{\rm h}\sim7.7$\,pc). They are rather more similar to the MW OH GCs ($\bar r_{\rm h}\sim$3.5\,pc), as evidenced by the K-S test probabilities listed in Table\,\ref{k-s-results}. In Sect.\,\ref{rh-discuss} we discuss the implication of this comparison.

It is worth noting, that in the SMC, a dIrr galaxy which closely resembles in total luminosity ($M_{V, SMC}=-16.82$\,mag) the galaxies in the current study, there exists only one old GC \cite[10.6\,Gyr, ][]{Dolphin01} with $r_{\rm h}\simeq5.6$\,pc.

Probability K-S tests give 1\% and 8\% likelihood that the bGC have a similar $r_{\rm h}$ distribution as the YHs or the old LMC GCs, respectively. The $r_{\rm h}$ distribution of LMC GCs is 32\% and 7\% similar to the ones of YHs and OHs, respectively (see Table\,\ref{k-s-results}).

\begin{figure}
\epsfig{file=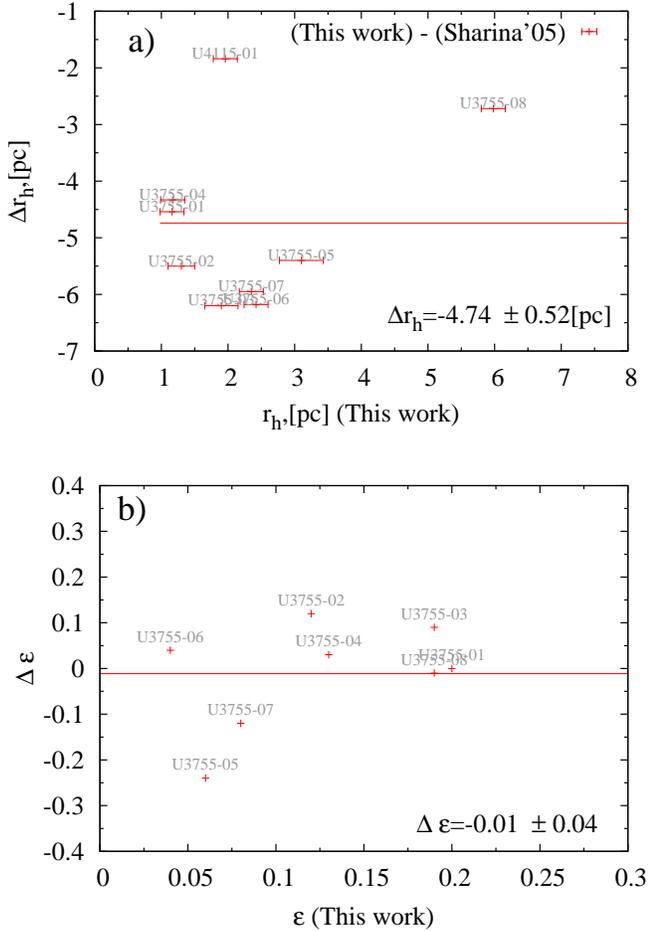, width=0.5\textwidth}
\caption{Differences in ${\rm r}_h$ (Panel a) and $\epsilon$ (Panel b) for our GCs in common with \protect\cite{Sharina05}. The least-square fit to the data (solid line) shows that our sizes are $\sim5$\,pc smaller than found in the earlier study, while the expected difference due to a different distance modulus is only $-0.55$\,pc.\label{struct_diff}}
\end{figure}

GCs more extended than those of the Milky Way and elliptical galaxies have been previously observed in other dIrrs \cite[]{Seth04, Sharina05}. We have nine GCs in common (one in UGC\,4115 and eight in UGC\,3755) with the \cite{Sharina05} study. Our $r_{\rm h}$ measurements, which take into account the cluster ellipticity, are significantly smaller by $\sim5$\,pc (see Fig.\,\ref{struct_diff}).
The expected difference due to updated distances to UGC\,3755 \cite[m--M$=29.35$,][]{Tully06} between both studies is just $-0.53$\,pc \cite[m--M$=28.59$,][adopted by \protect\cite{Sharina05}]{Karachentsev04}.

However, the absolute $r_{\rm h}$ comparisons between GCs should be taken with caution as they were assessed using different measurement methods by different studies. The structural parameters of the Galactic and LMC GCs are based on King model fits from \cite{McLaughlin&vdMarel05}. The same King model was used in our study, but using discrete concentration parameters as provided by {\sc ishape}, while \cite{McLaughlin&vdMarel05} derived a concentration parameter coming from their best $\chi^{2}$ fit to the model.

\subsubsection{Ellipticities}\label{ellipticities}

Two major dynamical effects are considered as drivers of the evolution of the GC ellipticities ($\epsilon$): {\it i)} the external galactic tidal field and/or {\it ii)} internal (age-dependent) rotation and/or velocity anisotropy \cite[e.g.][]{Fall&Frenk85,Han&Ryden94,Meylan&Heggie97}. A third possibility, although perhaps less attractive than general mechanisms, is that highly flattened GCs might arise from cluster mergers \cite[e.g.][]{deOliveira00}. \cite{Fall&Frenk85} showed that the cluster $\epsilon$ decreases by a factor $\sim 2-5\times t_{rh}$. The relaxation time depends on the clusters mass at $r=r_{\rm h}$. For Galactic GCs it ranges from $\sim10^{8}$ to $\sim10^{10}$\,yr \cite[]{Djorgovski93} and $\sim10^{8}-\sim10^{9}$\,yr for LMC GCs \cite[]{Fall&Frenk85}. It is well known, that on average the Magellanic Clouds GCs are flatter than those in the Milky Way and M31 \cite[e.g.][]{Geisler&Hodge80} and these differences might be primarily due to differences in cluster ages \cite[]{Han&Ryden94}.

The ellipticities of the clusters in our sample are measured at the $r_{\rm h}$ along the semi-major axis returned by {\sc ishape}.
\begin{figure*}
\epsfig{file= 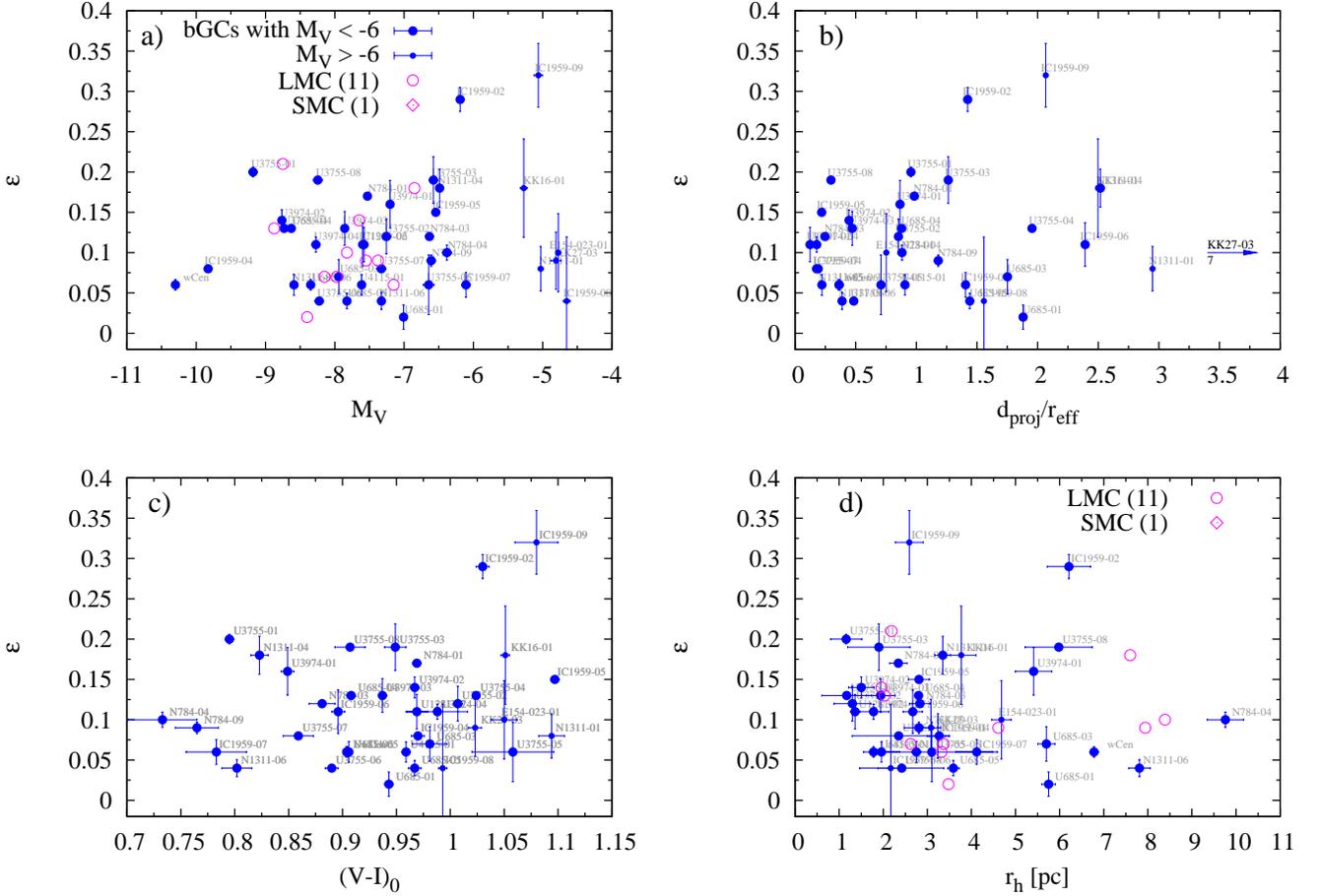,width=1\textwidth}
\caption{Correlation of clusters ellipticities with $M_V, d_{\rm proj}/r_{\rm eff}, (V-I)_0$ and $r_{\rm h}$ in panels a,\,b,\,c and d, respectively. Old GCs in the LMC and SMC are shown with open circles and diamonds, respectively \protect\cite[data from][]{Frenk&Fall82, Kontizas89,Kontizas90}.\label{ell}}
\end{figure*}
In Figure\,\ref{ell} we present the ellipticities of the bGCs in our dIrr galaxies versus $M_V, d_{\rm proj}/r_{\rm eff}, (V-I)_0$ and $r_{\rm h}$ in panels a,b,c and d, respectively. For comparison we show the old GCs in the Magellanic Clouds with data available from the literature \cite[]{Frenk&Fall82, Kontizas89,Kontizas90}. In general, the old LMC/SMC GCs overlap well with the clusters in our sample. As previous studies for various galaxy types also showed, no correlation between $\epsilon$ and the projected distance from the galactic center can be seen (Fig.\,\ref{ell}\,b).

\begin{figure}
\epsfig{file= 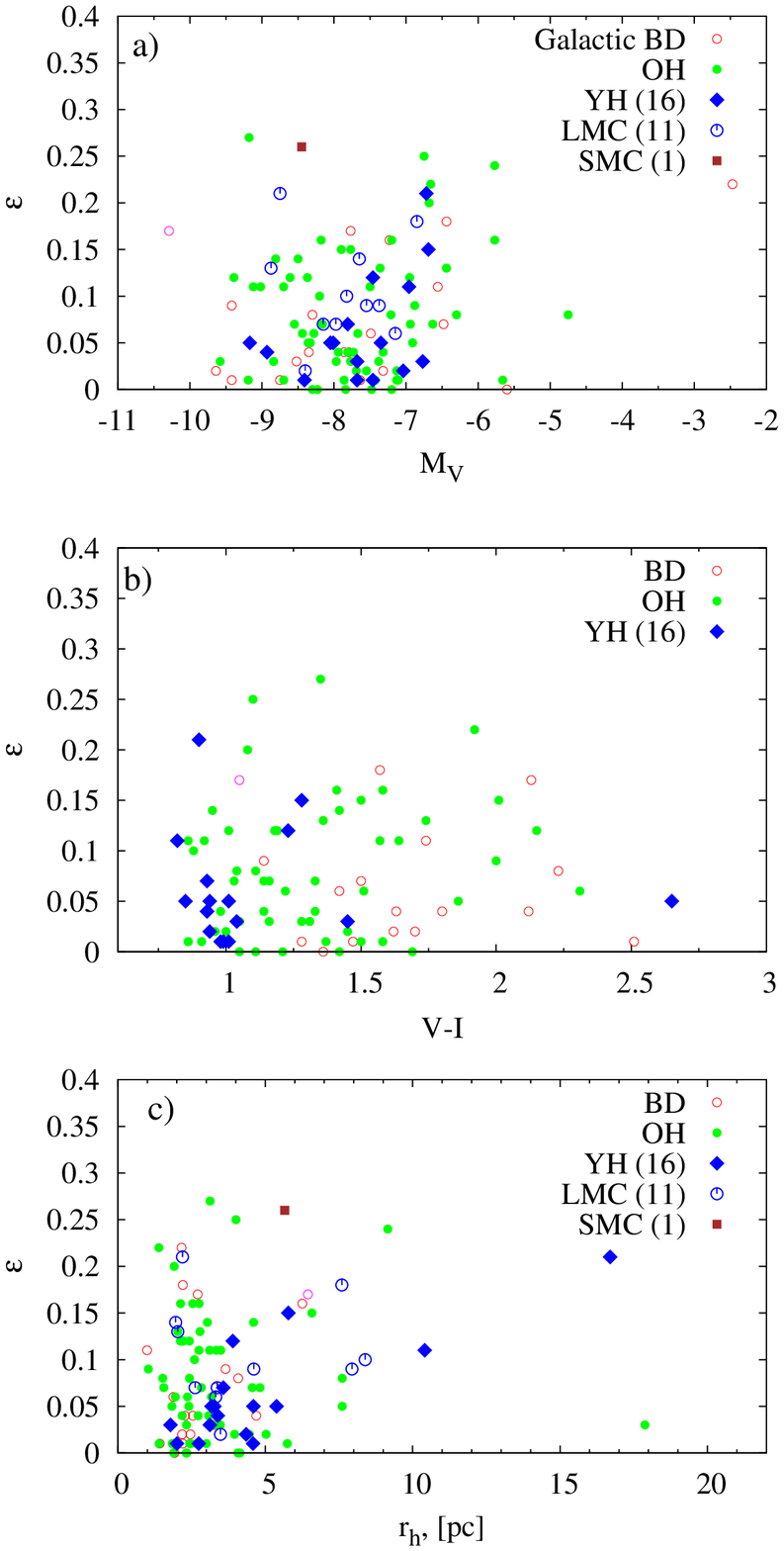,width=.5\textwidth}
\caption{Galactic and Magellanic Clouds GC ellipticities versus their a) $M_V$, b) $(V-I)_0$ and c) $r_{\rm h}$. Different symbols show the different MW and LMC/SMC GC populations. Data from the 2003 update of the \protect\cite{Harris96} MW GC catalog and from \protect\cite{Frenk&Fall82, Kontizas89,Kontizas90} for LMC and SMC.\label{mwgc_ell}}
\end{figure}

\begin{figure}
\epsfig{file=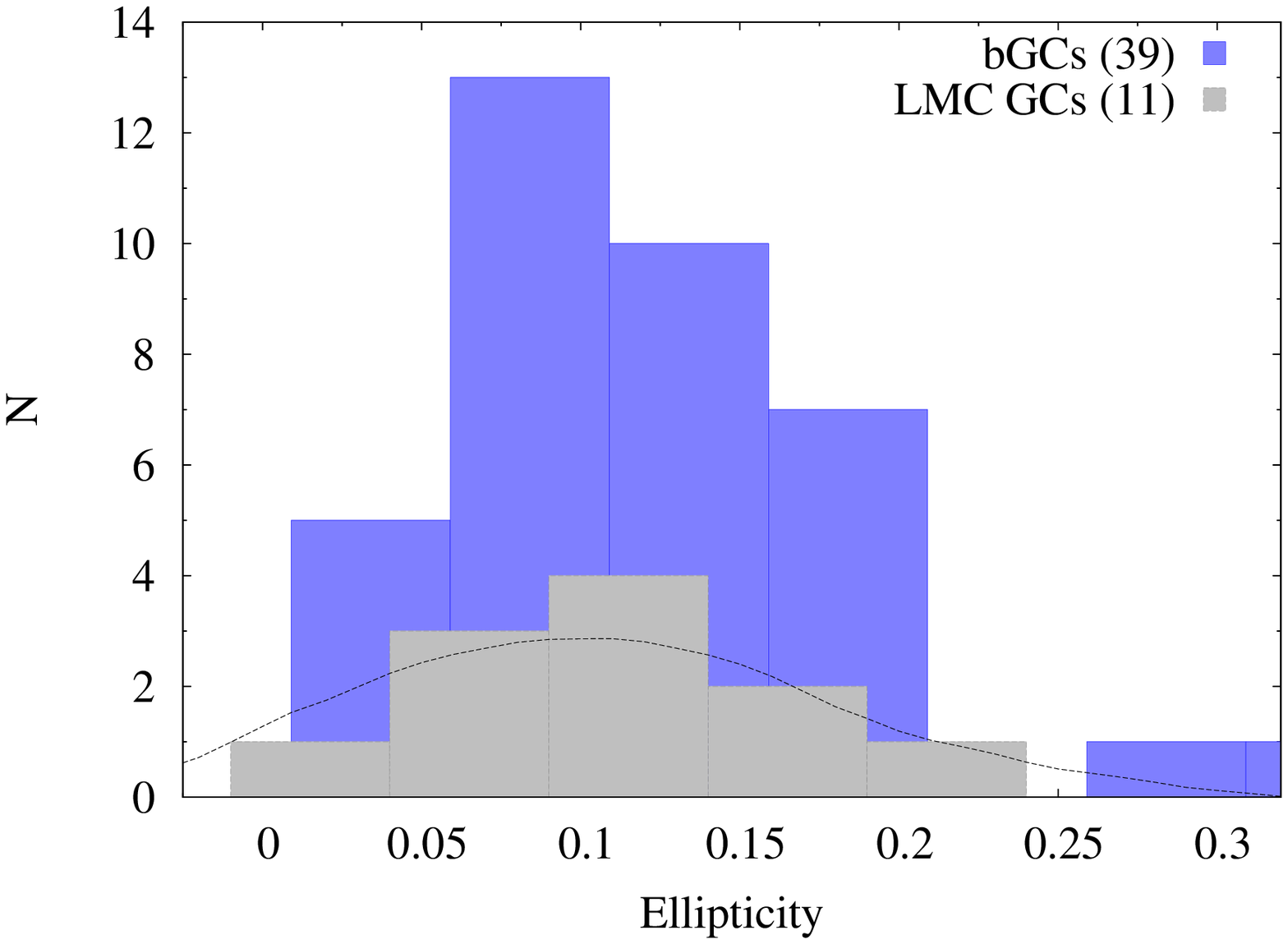,width=.5\textwidth}
\epsfig{file=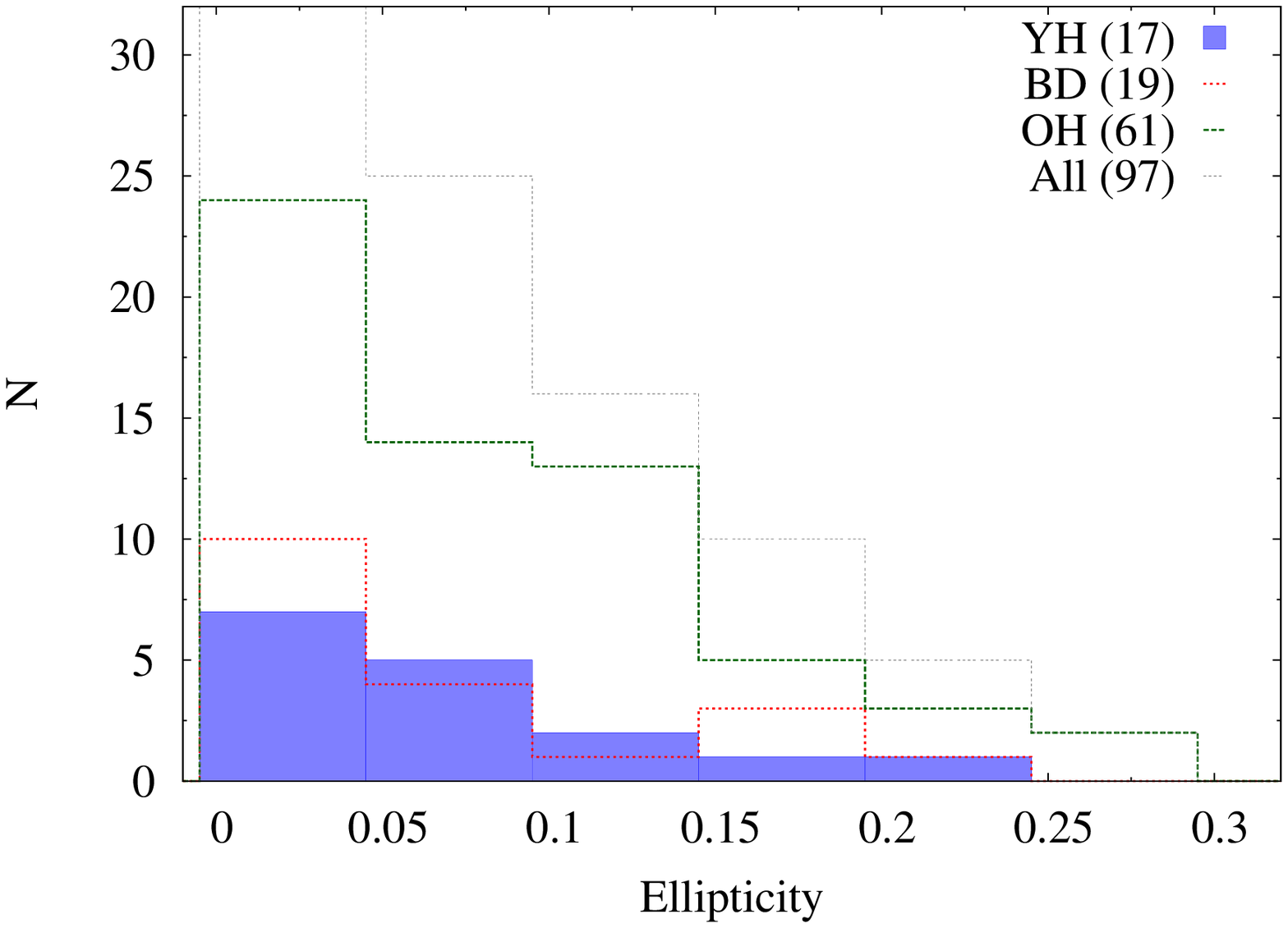,width=.5\textwidth}
\caption{Distributions of ellipticities for bGCs in our sample and old GCs in the LMC (upper panel) \protect\cite[]{Frenk&Fall82, Kontizas89} and in our Galaxy \protect\cite[lower panel, data from the 2003 update of the][catalog]{Harris96}.\label{ell_hist}}
\end{figure}

The comparison between the ellipticity vs. $M_V$, $(V-I)_0$ and $r_{\rm h}$ distributions the bGCs in this study with those of YH GCs in our Galaxy (suspected of being accreted from dwarf galaxies), is shown in Figure\,\ref{mwgc_ell}.
For the Galactic clusters, we used the $\epsilon$ entries in the 2003 update of the \cite{Harris96} catalog. Although only 16 YH GCs are included, it is seen that they overlap with the old Magellanic GCs in the $\epsilon$ vs. $M_V$ and $r_{\rm h}$ planes (Fig.\,\ref{mwgc_ell}\,a,\,c). However, the $\epsilon$ distributions (Fig.\,\ref{ell_hist}) of our bGCs and the LMC GCs are very similar to one another ($\bar\epsilon\simeq0.12\pm0.07$ and $\bar\epsilon\simeq0.11\pm0.06$) and flatter than that of the YH GCs ($\bar\epsilon\simeq0.06\pm0.06$), and show a peaked, Gaussian-type distribution.
The result of a Kolmogorov--Smirnov test (see Table\,\ref{k-s-results}) run on the $\epsilon$ distributions of the YH GCs, LMC old GCs and the bGCs in our sample showed that there is only 1 and 3\% chance that the LMC old GCs or bGCs, respectively, are drawn from the same distribution as the YH GCs, while the bGCs and the LMC GCs share the same distribution at a 99\% confidence level. The presence of such a turnover might mean that either the clusters are not fully relaxed, i.e. they are dynamically young and still evolving, or it is an indication of the impact of the weaker potential of these dwarfs on the initial conditions during the GC formation, or both.

Therefore, if the YH GCs are of external origin, their $\epsilon$ distribution should be similar to that of the bGCs. In contrast, it is more like a power law as seen in the Fig.\,\ref{ell_hist}, lower panel. This might be an indication that their accretion must have happened very early and the strong tidal field of our Galaxy might be responsible for the evolution of the $\epsilon$ distribution away from the initial conditions, removing the turnover. Note, however, that reliable ellipticities are available only for approximately half of all MW YH GCs (16 out of 30) and for old GCs in the LMC (11 out of 17). Thus the ellipticity distribution we show might be biased.

\subsection{Specific frequencies}\label{SN}

The 'specific frequency' ($S_{N}$) is a quantity introduced by \cite{Harris&vdBergh81} to compare the richness of GCSs between (elliptical) galaxies. The $S_{N}$ was originally defined as twice the number of GCs down to the turnover peak of the GC luminosity function, or (as most commonly used) as the total number of GCs ($N_{\rm{GC}}$), normalized to a galaxy luminosity of $M_{V}=-15$\,mag:
\begin{equation}
\hspace{2.5cm}S_N = N_{\rm{GC}}10^{0.4(M_{V}+15)}
\end{equation}

For poor GCSs containing few GCs, such as those in our dIrrs, calculating the $S_N$ as originally defined is practically impossible. Hence, the $S_N$ values presented here were derived from the total $N_{\rm{GC}}$. Given the small physical sizes of the studied dIrrs and the large enough ACS field of view, there is no need to apply a geometrical completeness correction to the total GC number for most of the galaxies except for the brightest and largest two, NGC\,784 and ESO\,154-023. Because of the extremely deep ACS observations, corrections due to photometric incompleteness are not necessary either (see Sect.\,\ref{comcon}). In order to derive the total galaxy light we used the IRAF/STSDAS task {\sc ellipse}. The galaxy magnitudes we derived (see Table\,\ref{main}) were identical (within the errors) to the RC3 \cite[]{deVaucouleurs91} values for the same galaxies at the radius of the $\mu_{B}=25$\,mag arcsec$^{-2}$ isophote. However, we could not directly measure the total magnitudes for NGC\,784 and ESO\,154-023 which extend beyond the ACS field and thus we adopted their RC3 magnitudes.

The majority of the bGCs in our ACS images are resolved and the estimated background contamination for this GC sub-population is low (see Sect.\,\ref{comcon}). Thus, the uncertainty of the $S_N$ values for the bGCs presented in Table\,\ref{SNTable} is mainly due to stochasticity in the number of genuine GCs rather than background contamination. The subpopulation of red GCs is more likely affected by faint and unresolved background sources, because red GCs are on average fainter and more compact than bGCs. This is the reason why our analysis is mainly focused on bGCs.

\subsubsection{Present-day $S_N$ Values}\label{SNp}

Using the distance modulus for each galaxy as determined by \cite{Tully06} the apparent galaxy magnitudes were converted to absolute ones. Taking the numbers of GCs as given in Table\,\ref{SNTable}, the resulting $S_N$ values then range from 0.74 to 88. Excluding the extreme outlier (KK\,27) with $S_N=88$, the mean value for all galaxies with blue GCs is $\sim3.8$. The total mean $S_N$ would increase to $\sim4.2$ if one would include the red GCs.
\begin{figure}
\epsfig{file= 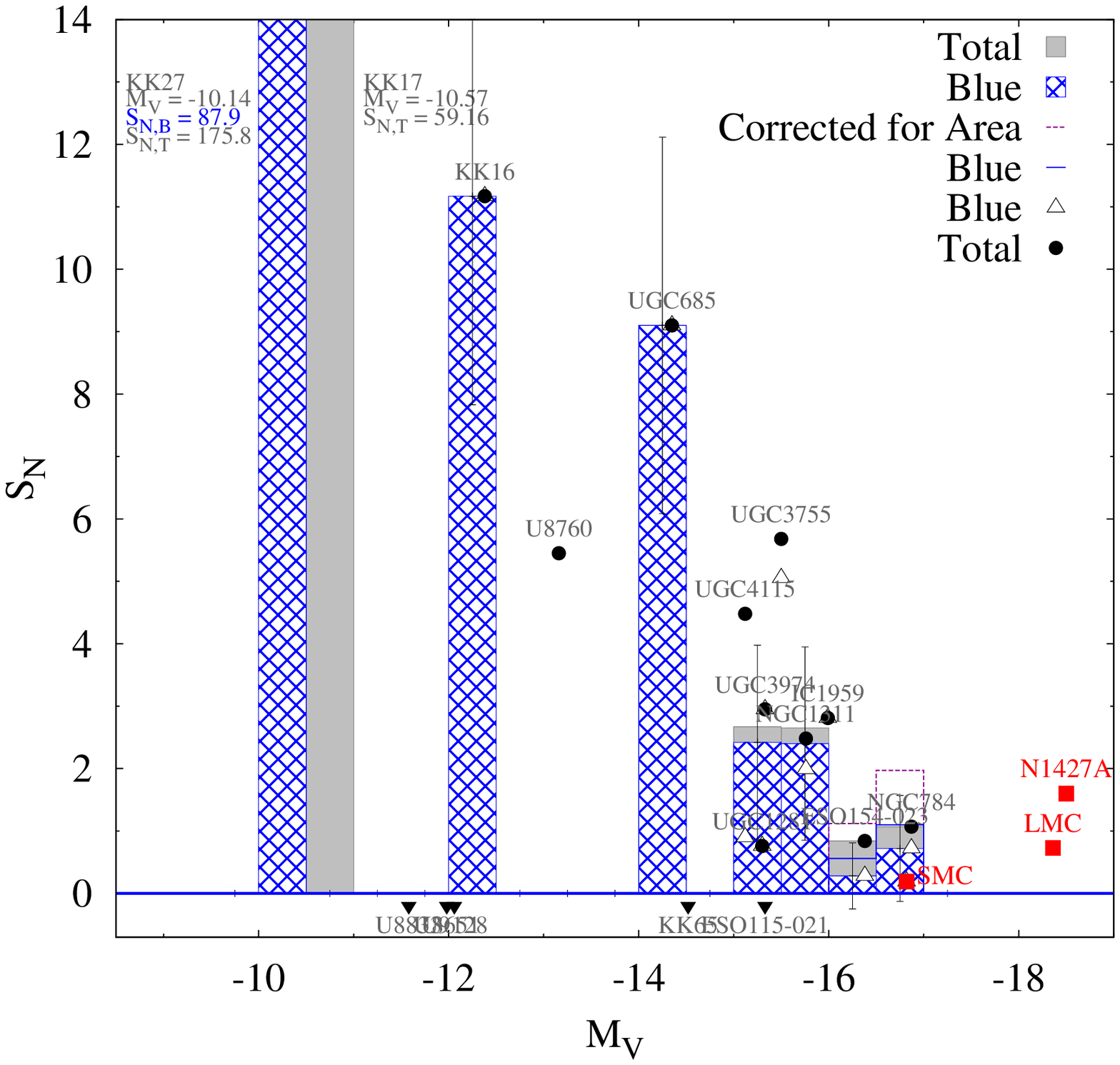,width=0.5\textwidth}
\caption{$S_N$ versus host galaxy $M_V$. Open triangles and filled circles are the $S_N$ values for our sample dIrrs including only the blue GCs and blue and red GCs for each individual galaxy, respectively. Galaxies in which no GC candidates were found are shown with filled triangles and negative $S_N$. With the hatched histogram we show the average $S_N$ distribution per galaxy magnitude bin with width equal to the width of the histogram bin. For comparison with filled squares are shown the $S_N$ values for the Magellanic Clouds and NGC\,1427A in the Fornax galaxy cluster.\label{SNfig}}
\end{figure}
Figure\,\ref{SNfig} shows the $S_N$ values for each individual galaxy (symbols) and an average $S_N$ for galaxies in magnitude bins (histogram). The galaxies in which no GC was detected are plotted with negative $S_N$ values. The $S_N$ values of the Magellanic Clouds and NGC\,1427A are shown for comparison. To calculate the Magellanic Clouds' absolute magnitudes we used the distance moduli of \cite{Laney&Stobie94} and information for NGC\,1427A was taken from \cite{Georgiev06}. Note that the dIrrs in our sample with luminosities similar to the SMC also have similar $S_N$ values. In general we find that the GC specific frequency in dIrrs increases (on average) with decreasing host galaxy luminosity. This result was also found for early-type dwarfs in galaxy clusters \citep{Miller98, Forbes05, Strader06,Miller&Lotz07}.

\subsubsection{$S_N$ Values After Evolutionary Fading}\label{SNf}

Since the dIrr galaxies in our sample are of Magellanic type with ongoing star formation, their present-day integrated luminosity is dominated by the young stellar populations that formed in the most recent star formation episode. In order to make a more relevant comparison with the $S_N$ values in evolved dwarf galaxies like dEs or dSphs, we have to account for fading of the dIrrs' luminosity due to stellar evolution. All galaxies in our sample have luminosities comparable to or even fainter than the SMC. We therefore adopted the SMC metallicity of [Fe/H]$=-0.68$ \cite[]{Luck98} for the passive evolution of the stellar population. \cite{BC03} SSP models were used to derive the stellar age for each galaxy based on their color and adopted metallicity. The mean stellar age for all galaxies as inferred by their $V-I$ colors is 1.3\,Gyr, with individual ages ranging from 0.4 to 3.7\,Gyr. Given the low surface brightness and low star formation of these dwarfs, a passive (dissipationless) evolution was assumed to evolve the galaxies' luminosity to an age of 14\,Gyr. This resulted in an average luminosity fading of $2.34\pm0.43$ mag, with a full range from 2.98 to 1.09\,mag. This leads to an increase in the $S_N$ values by a factor of 2 to almost 16 due to passive evolutionary fading. The specific frequencies after age fading ($S_{Nf}$) are listed in Table\,\ref{SNTable}.

It is worth assessing the impact of the adopted SMC metallicity on the evolved $S_N$ values. Due to the continuous star formation rate, typical for late-type galaxies, a dIrr galaxy in our sample cannot be more metal-poor than its most metal-poor GC(s); unless, in the unlikely scenario where the galaxy underwent instantaneous star formation which ceased abruptly prior to GC formation. However, assuming a mean galaxy metallicity as metal-poor as $0.02\times Z_{\odot}$, the mean stellar age, as inferred from the measured $V-I$ color, increases by $\sim1$\,Gyr. Hence, after evolutionary age fading the galaxy will appear, on average, $\sim1$\,mag fainter. This translates into an increase of $S_N$ up to 6 times of the present-day value, unlike up to 16 times if the dIrrs were of SMC metallicity. In general, the change of $S_N$ values due to evolutionary age fading will be milder for more metal-poor galaxies, and the evolved $S_N$ values in Table\,\ref{SNTable} will apply only to galaxies with SMC metallicity ([Fe/H$]=-0.68$, i.e. similar to $0.2\times Z_{\odot}$ in the \cite{BC03} SSPs). This illustrates the dependence of $S_N$ values on host galaxy metallicity.

If one would add all galaxies with GCs together, a master galaxy of $M_V=-18.24$\,mag would result with a $S_{Nb}= 1.9$. Including also dIrrs without GCs results in a master galaxy of $M_V=-18.36$\,mag and $S_{Nb}=1.7$. Fading the master galaxy light to an age of 14\,Gyr gives for the two cases $M_V=-16.0$ and $M_V=-16.11$\,mag and specific frequencies of $S_N=14.8$ and $S_N=13$, respectively. Again, this is consistent with $S_N$ values for dEs/dSphs of similar absolute magnitude.
 
\section{Discussion and conclusions}\label{conclusions}

We have searched for old globular clusters in 19 Magellanic-type dwarf Irregular (dIrr) galaxies utilizing archival $F606W$ and $F814W$ HST/ACS images. The main goal of this study is to assess the properties of the old GCs in such systems and compare them with those of the Galactic ``Young Halo'' GCs (YH GCs), which are considered to be of external origin \cite[e.g.][]{Zinn93, Mackey&vdBergh05, Lee07}. The studied dIrrs reside in nearby (2 - 8\,Mpc) groups and associations containing dwarf galaxies only, with no massive nearby galaxy around \cite[]{Tully06}. All galaxies in our sample have absolute magnitudes smaller than or equal to the SMC.

\subsection{An Apparent Lack of Low-Luminosity GCs}\label{lumcol-discuss}

We detect in total 50 GCs in 13 dIrrs, of which $37$ have $(V-I)$ colors consistent with ``blue'' (old and metal-poor) GCs (bGC) typical for low-mass dwarfs and the bGC population in other galaxies. The few red GCs (rGC) we detect are indistinguishable in sizes from the bGCs and are not expected in this galaxy type, perhaps they are background contaminants. Only follow-up spectroscopy can reveal their nature. The luminosity function (LF) of the bGC candidates in our sample shows a turnover magnitude at $M_V=-7.41\pm0.22$\,mag, similar to what is typically found in other galaxy types, and a GCLF width of $\sigma=1.79\pm0.31$, typical for dIrrs but slightly broader than the typical width of GCLFs in massive galaxies. The broadening of the GCLF might be (at least partly) due to the distance modulus uncertainties to those dwarfs of $\sigma(m-M)\simeq0.05$ \cite[]{Tully06}.

Interestingly, we do not find many low-luminosity GCs, contrary to expectations. As we showed, this is not due to incompleteness or selection criteria (Sect.\,\ref{magcol}). The lack of faint GCs might be related to environmental effects and the physical conditions of the GC formation and disruption in such dwarfs. However, all dIrrs in our sample are found in associations of dwarf galaxies, e.g. there is no dominant massive galaxy in their immediate vicinity. Thus tidal stripping or disruption in the current environment is unlikely to be responsible.

Another possibility could be that the bluer $(0.8\sim V-I)$ and slightly brighter $(M_V\sim-6.5$\,mag) GCs seen in Fig.\,\ref{cmd} might actually be relatively young GCs. Aging due to stellar evolution would make their magnitudes fainter and colors redder. If so, a shift in their colors and magnitudes of $(V-I)\sim0.15$ and $M_V\sim1.5$\,mag would be required to populate that region in the CMD. According to the \cite{BC03} SSP models, this is exactly the amount of color and magnitude evolution needed for a $\gtrsim4$\,Gyr old metal-poor cluster with $(V-I)\gtrsim0.8$, to evolve to $\sim14$\,Gyr.

Further detailed spectral analysis of GC candidates in dIrrs are necessary to address this question and draw firmer conclusions. Although an interesting observational result, for now we can not extend the discussion further given the inadequate number statistics in our data set.

\subsection{Sizes: GCs in dIrr Galaxies are smaller than ``Young Halo'' GCs}\label{rh-discuss}

The half-light radii, $r_{\rm h}$, of the bGCs in our sample are smaller on average, $r_{\rm h}\sim3.3$\,pc, than those of the YH GCs in our Galaxy ($r_{\rm h}\sim7.7$\,pc) and those in the Magellanic Clouds and Fornax dwarfs ($r_{\rm h}\sim6.3$\,pc) and other dIrrs \cite[]{Sharina05,vdBergh06}. They compare well with the $r_{\rm h}$ values of Galactic OH GCs, as shown by the K-S tests in Table\,\ref{k-s-results}. This suggests that either the conditions of YH GC formation were different from those in low-mass galaxies, or their sizes were influenced by tidal shocks in their subsequent evolution in the Galactic tidal field which will require that they have been accreted very early. $N-$body simulations \cite[e.g.][]{Dinescu99,Gieles07} show that the effect of bulge/disk shocks scales with $r_{\rm h}^3/M_{\rm cl}$. Hence, depending on their orbital periods and pericentric radii, the tidal field will strongly affect diffuse clusters, such as the YH GCs, and transform their $r_{\rm h}$ distributions. Further, \cite{Lamers05} show that the cluster disruption time scales with the ambient density. As $r_{\rm h}$ is stable over many relaxation times (Sect.\,\ref{rh}) in the absence of a tidal field, a modeling of the $r_{\rm h}$ evolution with the strength of the host galaxy tidal field and environmental density will be required to understand the $r_{\rm h}$ evolution.

\subsection{The Ellipticity Distribution of GCs in dIrr Galaxies}\label{ell-discuss}

We find that, on average, the ellipticities ($\epsilon$) for the GCs in the studied dIrrs are practically the same as those of the old GCs in the Magellanic Clouds ($\bar\epsilon\simeq0.12\pm0.07$ and $\bar\epsilon\simeq0.11\pm0.06$, respectively). They also show similar distributions of $\epsilon$ with $M_V$, $r_{\rm h}$ and cluster color (Fig.\,\ref{ell}). Although ellipticities are only available for half the Galactic YH GCs, their distribution differs significantly from those of GCs in our sample and those in the LMC (see Fig.\,\ref{ell_hist}). This is further shown by the result of the K-S tests presented in Table\,\ref{k-s-results}. This shows that {\it if\/} the YH GCs formed in similar environments as the GCs in low-mass dIrr galaxies, their ellipticity must have been significantly altered by the tidal field of the Galaxy after their accretion. I.e., if YH GCs were accreted from satellite galaxies, the accretion event(s) must have happened quite early.

The observed turnover in the ellipticity distributions for the bGCs and LMC old GCs in contrast with the power-law $\epsilon$ distribution for the Galactic YHs, might suggest that the former are slightly younger and not yet fully relaxed (see also Sect.\,\ref{ellipticities}), as expected from theory \cite[]{Fall&Frenk85}.

Furthermore, the strength of the galactic tidal field is expected to be important in setting the initial conditions during the GC formation. It probably also causes the mean $\epsilon$ value to vary with galaxy mass, as slightly higher $\epsilon$ values are observed for the young clusters in the SMC than in the LMC \cite[]{Kontizas90}. Thus, it does not seem surprising that the GCs in our low-mass dIrrs and in the Magellanic Clouds on average are flatter than those in massive galaxies and that, if the YH GCs were accreted early enough, their evolution in the stronger Galactic tidal field must have influenced their ellipticities. The fact that the $\epsilon$\,vs.\,$r_{\rm h}$ relation for the YH GCs holds just as well for GCs in our dIrrs and the LMC/SMC suggests that they might all have formed in similar initial conditions. However, this interpretation needs to be tested with larger numbers of GCs in dIrrs, simulations and further precise determinations of the GC ellipticities in various environments, as well as future integrated spectra to infer their ages and metallicities.

\subsection{Specific Frequencies: Trend with galaxy luminosity and Comparison to Dwarf Ellipticals}\label{SN-discuss} 

The present-day specific frequencies ($S_N$) of the dIrr galaxies in our sample span a broad range from $S_N=0.3$ to 87. Using the galaxy $V-I$ colors and the SSP models of \citet{BC03} with the metallicity of the SMC, we derive age fading of the galaxies' luminosity to 14\,Gyr. We find that this causes an increase in the $S_N$ by a factor of $\sim2$ to 16, if all GCs survive and no new ones are formed (see also Table\,\ref{SNTable}).

Assuming that these groups of dIrr galaxies might not be fully relaxed yet \cite[]{Tully06} and that their members will merge eventually, we constructed a ``master'' dIrr galaxy from all members. The resulting present-day and faded $S_N$ values of the ``master'' dIrr are similar to the values for early-type dwarfs. This supports the idea that some dE/dSphs might have evolved from dIrrs \cite[]{Miller98}, due to gas loss after the main burst of star formation \cite[]{Dekel&Silk86} and/or via external ram pressure stripping \cite[]{Moore98,Grebel03}. The $S_{N}$ values of the non-faded and faded dIrrs, in general, are consistent with those found for dEs in the same luminosity range \cite[]{Miller98,Miller&Lotz07}.

The recent discovery of hidden non-axisymmetric disk-like structures and other morphological anomalies in Virgo early-type dwarfs from detailed isophotal analysis \cite[]{Ferrarese06} also suggests that their progenitors might be late-type dwarfs. This view is theoretically supported by \cite{Mastropietro05}. If rapid cessation of star formation processes occurs in present-day dIrrs, the comparison with \cite{BC03} stellar evolution models shows that $\sim10$\,Gyr of passive fading are required for dIrrs to fade and fall on the dSphs metallicity--luminosity relation \cite[see also][]{Jerjen04}.

If all dIrrs in our sample are merged together (without dissipation) to a master dIrr galaxy and the galaxy light is evolved to 14\,Gyr, the GCS characteristics (Table\,\ref{SNTable}) of the ``merger remnant'' resemble those of a field dwarf elliptical or spheroidal galaxy as suggested by \cite{Miller98}.

\section{Summary and Outlook}

Our study shows that properties of old GCs in low-mass dIrr galaxies hold interesting and important insights in our understanding of what were the conditions at the time of their formation. Indeed, considering their color and magnitude distributions, they are as old and as metal poor as the GCs in other galaxy types. Their structural parameters (${\rm r}_h$ and $\epsilon$), whose evolution is shown to depend mainly on processes internal to the GCs (e.g. rotation, two-body relaxation), are very similar to those of the old GCs in the Magellanic Clouds, but differ from YH GCs in our Galaxy. Under the assumption that initial conditions of GC formation (occurring in giant molecular clouds) were similar among old GCs, this difference suggests that the initial structural parameters were altered by an amount depending on the environment, i.e., the strength of the galactic potential plus external pressure. Therefore, our comparison of the bGCs in dIrrs with the Galactic YH GCs, the latter of which are suspected to have their origin in accreted low-mass galaxies, may indicate that the Galactic YH GCs were accreted very early. This accretion must have been early enough for the Galactic tidal field to have an impact on the time scale of the evolution of the structural parameters.

However, to properly assess all the addressed suggestions to a higher statistical confidence, we suggest that further detailed studies of GC properties in other samples of low-mass dIrr galaxies be undertaken in the near future.

%===== Acknowledgments =====

\acknowledgments

{\footnotesize ACKNOWLEDGMENTS}\\

IG is grateful for the award of a STScI Graduate Research Fellowship. Support for this work was provided in part by NASA through HST grant number GO-10550 from the Space Telescope Science Institute, which is operated by the Association of Universities for Research in Astronomy, Inc., under NASA  
contract NAS5-26555. THP acknowledges support through a Plaskett Fellowship at HIA. MH acknowledges support from a German Science Foundation Grant (DFG-Projekt HI-855/2). The authors are grateful to the anonymous referee whose comments helped to improve this work. We would like also to thank Prof. Klaas de Boer for his valuable suggestions which improved the text.

{\it Facilities:} \facility{HST (ACS)}

%===== References =====
\bibliographystyle{apj}
\bibliography{references}

\begin{center}
\begin{table*}
\footnotesize
\caption{General properties of the studied dwarf irregular galaxies. In column
  (1) are listed the Group ID to which the galaxy belongs and the galaxy's
  name; columns (2) and (3) list the galaxies' coordinates, columns (4) and
  (5) their morphological classification; columns (6) and (7) give the
  distance and distance modulus from \protect\cite{Tully06}, followed by the
  galaxies'  
apparent (8) and absolute (9) magnitudes and colors (10) measured in this work
(all dereddened from Galactic extinction).\label{main}} 
%\hspace{-.8cm}
\begin{tabular}{lp{1.5cm}p{1.5cm}ccccccc}
\tableline\tableline
Group/ID & R.A. & Decl. & \multicolumn{2}{c}{Morph. Type} & D\tablenotemark{a} & $m-M$\tablenotemark{a} & $V_0$ & $M^{0}_{V}$ & $(V-I)_0$ \\
& (J2000.0) & (J2000.0) & \multicolumn{2}{c}{} & Mpc & mag & mag & mag & mag \\
             &	 	  &	 	&           &         &        &         &        &         &         \\
(1)        &	(2)	  &	(3)	&  (4)    &   (5) &  (6) &  (7)  & (8)  & (9)   & (10) \\
             &	 	  &	 	&           &         &        &         &        &         &         \\
\tableline
14+12        &		  &		&      &       &      &       &       &        & \\
ANTLIA       & 10 04 03.9 & $-$27 20 01 & 10.0 &  IAB  & 1.25 & 25.49 & 18.39 & -7.1   & 0.80 \\
\tableline
14+08        &		  &		&      &       &      &       &       &        &  \\
UGC\,8760    & 13 50 50.6 & $+$38 01 09 & 9.8  &  IB   & 3.24 & 27.55 & 14.39 & -13.16 & 0.68 \\
UGC\,8651    & 13 39 53.8 & $+$40 44 21 & 9.9  &  I    & 3.02 & 27.40 & 15.42 & -11.98 & 0.66  \\
UGC\,8833    & 13 54 48.7 & $+$35 50 15 & 9.9  &  IAB  & 3.20 & 27.53 & 15.95 & -11.58 & 0.58 \\
\tableline
17+06\tablenotemark{b} &      &       &      &       &       &        &       &   & \\
NGC\,784     & 02 01 17.0 & $+$28 50 15 & 7.7  &  SBdm & 5.19 & 28.58 & 11.71 & -16.87\tablenotemark{c} & -- \\
UGC\,1281    & 01 49 31.5 & $+$32 35 17 & 7.5  &  Sdm  & 5.13 & 28.55 & 13.25 & -15.30 & 0.83 \\
KK\,16       & 01 55 20.3 & $+$27 57 14 & 10.0 &  I    & 5.47 & 28.69 & 16.31 & -12.38 & 0.88 \\
KK\,17       & 02 00 10.2 & $+$28 49 53 & 10.0 &  I    & 4.91 & 28.45 & 17.88 & -10.57 & 0.96 \\
UGC\,685     & 01 07 22.4 & $+$16 41 02 & 9.1  &  Sm   & 4.70 & 28.36 & 14.01 & -14.35 & 0.82 \\
\tableline
14-14        &		  &		&      &       &      &       &       &        &    \\
ESO\,115-021 & 02 37 48.1 & $-$61 20 18 & 7.6  &  SBd  & 4.99 & 28.49 & 13.16 & -15.33 & 0.72 \\
KK\,27       & 03 21 02.4 & $-$66 19 09 & 10.0 &  I    & 4.15 & 28.09 & 17.95 & -10.14 & 0.78 \\
\tableline
14+14        &		  &		&      &       &      &       &       &        &    \\
ESO\,154-023 & 02 56 51.2 & $-$54 34 23 & 8.8  &  SBm  & 5.76 & 28.80 & 12.42 & -16.38\tablenotemark{c} & 0.680\tablenotemark{c} \\
IC\,1959     & 03 33 09.0 & $-$50 24 42 & 8.6  &  SBm  & 6.06 & 28.91 & 12.92 & -15.99 & 0.74 \\
NGC\,1311    & 03 20 06.7 & $-$52 11 13 & 8.8  &  SBm  & 5.45 & 28.68 & 12.92 & -15.76 & 0.78  \\
\tableline
14+19        &		  &		&      &       &      &       &       &        &     \\
UGC\,3974    & 07 41 52.0 & $+$16 47 54 & 9.8  &  IBm  & 8.04 & 29.53 & 14.20 & -15.33 & 0.98 \\
UGC\,4115    & 07 57 02.4 & $+$14 23 01 & 9.9  &  I    & 7.72 & 29.44 & 14.32 & -15.12 & 0.74 \\
KK\,65       & 07 42 32.0 & $+$16 33 39 & 10.0 &  I    & 7.72 & 29.52 & 15.00 & -14.52 & 0.76 \\
UGC\,3755    & 07 13 51.6 & $+$10 31 19 & 9.9  &  IAB  & 7.41 & 29.35 & 13.85 & -15.50 & 0.88 \\
\tableline
Dregs        &		  &		&      &       &      &       &       &        &    \\
UGC\,9128    & 14 15 56.5 & $+$23 03 19 & 9.9  &  IAB  & 2.24 & 26.75 & 14.69 & -12.06 & 0.60 \\
\tableline\tableline
\end{tabular}
\tablenotetext{a}{\protect\cite{Tully06}}
\tablenotetext{b}{Information for UGC\,9240, which belongs to this group, was taken from \protect\cite{Sharina05}}
\tablenotetext{c}{From RC3}
\end{table*} 
\end{center}

\clearpage

\begin{table*}
\begin{minipage}{160mm}
\scriptsize
\caption{Specific frequencies. In column (1) are listed the Group ID to which the galaxy belongs and the galaxy's name; The following columns are: the distance (2), absolute magnitude (3), number of blue GCs (4), number of red GCCs (5), blue GCs and total specific frequency $S_{Nb} (S_{Ntot})$ (6), age of the dominant stellar population deduced from the \protect\cite{BC03} SSP ($0.2\times Z_{\odot}$) using the galaxy color (7), amount of $V-$band fading to 14\,Gyr (8), $M_{V,f}$ after fading (9), $S^{f}_N$ after fading (10), $M_V$ of the product galaxy after merging all group members together (11), $S_N$ of the merger product, amount of the merger product fading to 14\,Gyr (12), $M_{V {\rm grp},f}$ after merger fading (13), $S^{f}_{N}$ after merger fading (14)\label{SNTable}. Numbers in brackets always refer to the total values.}
%\hspace{-.8cm}
\begin{tabular}{p{2cm}p{.4cm}cp{.1cm}p{.1cm}p{1cm}p{.6cm}p{.7cm}p{1cm}p{1cm}p{1cm}p{.8cm}p{.5cm}p{.8cm}p{.4cm}}
\tableline\tableline
Group/ID &D\tablenotemark{a} & $M^{0}_{V}$ & $N_{b}$ & $N_{r}$ & $S_{N{\rm b (tot)}}$ & Age & Fade & $M^{0}_{V,f}$ & $S^{f}_{N{\rm b (tot)}}$ & $M^{0}_{Vgrp}$ & $S_{N{\rm b (tot)}}$ & Fade & $M^{0}_{V {\rm grp},f}$ & $S^{f}_{N{\rm b (tot)}}$ \\
 & Mpc & mag & & & & Gyr & $\Delta$\,mag & mag & & mag & & $\Delta$\,mag & mag & \\
             &        &         &         &          &           &         &         &          &          &         &         &         &          &    \\
(1)        &	(2)	  &	(3)	&  (4)    &   (5) &  (6) &  (7)  & (8)  & (9)   & (10) & (11) & (12)  &  (13)  & (14) & (15) \\
             &        &         &       &         &           &         &         &          &          &         &         &         &          &    \\
\tableline
14+12        &      &       &       &        &       &             &    &   &     &   --   & -- &  --  &   --  & -- \\
ANTLIA      & 1.25 & -7.1   & 0 & 0 & 0      (0)  & 1.2 & 2.36 & -4.74 &0   (0) \\
\tableline
14+08        &      &       &       &        &       &             &   &    &    & -14.82 & 0 (0) & 2.72 & -12.1 & 0 (0) \\
UGC\,8760    & 3.24 & -13.16 & 0 & 1 & 0 (5.45) & 0.8 & 2.57 & -10.59 & 0   (58) \\
UGC\,8651    & 3.02 & -11.98 & 0 & 0 & 0      (0)  & 0.7 & 2.62 & -9.36 & 0 (0)  \\
UGC\,8833    & 3.20 & -11.58 & 0 & 0 & 0      (0)  & 0.4 & 2.98 & -8.60 & 0 (0) \\
\tableline
17+06\tablenotemark{b} &       &        &       &      &       &       &        &       &        & -17.20 & 1.45 (1.85) & 2.18 & -15.02 & 10.78 (13.72) \\
NGC\,784    & 5.19 & -16.87\tablenotemark{c} & 4 & 2 & 0.72 (1.07) & 1.0 & 2.36 & -14.51 & 6.3 (9.4) \\
UGC\,1281  & 5.13 & -15.30 & 1 & 0 & 0.76 (0.76) & 1.9 & 2.37 & -12.93 & 6.7  (6.7) \\
KK\,16      & 5.47 & -12.38 & 1 & 0 & 11.2 (11.2)& 1.2 & 2.42 & -9.96 & 104 (104) \\
KK\,17      & 4.91 & -10.57 & 0 & 1 & 0 (59.2)& 2.9 & 1.39 & -9.18 &   0   (213) \\
UGC\,685& 4.70 & -14.35 & 5 & 0 & 9.10 (9.10) & 1.9 & 2.37 & -11.98 & 81 (81) \\
\tableline
14-14        &      &       &       &        &       &      &       &        &     & -- &  --  &   --  & -- & -- \\
ESO\,115-021& 4.99 &-15.33 & 0 & 0 &  0   (0)  & 1.1 & 2.33 & -13.0 &   0   (0) \\
KK\,27      & 4.15 & -10.14 & 1 & 1 & 88 (176) & 1.2 & 2.35 & -7.79 & 766 (1531) \\
\tableline
14+14      &      &       &       &        &       &      &       &       &        & -17.27 & 1.49 (1.86) & 2.41 & -14.86 & 13.69 (17.11) \\
ESO\,154-023& 5.76 & -16.38\tablenotemark{c} & 1 & 2 & 0.28 (0.84) & 0.8 & 2.55 & -13.83 & 2.9 (8.8) \\
IC\,1959    & 6.06 & -15.99 & 7 & 0 & 2.81 (2.81) & 1.1 & 2.33 & -13.66 & 24.1 (24.1) \\
NGC\,1311& 5.45 & -15.76 & 4 & 1 & 1.99 (2.48) & 1.2 & 2.35 & -13.41 & 17.3 (21.6) \\
\tableline
14+19       &      &       &       &        &       &            &     &     &     & -16.68 & 2.8 (3.8) & 2.05 & -14.63 & 18.29 (25.31) \\
UGC\,3974& 8.04 & -15.33 & 4 & 0 & 2.95 (2.95) & 3.7 & 1.09 & -14.24 & 8.1  (8.1) \\
UGC\,4115& 7.72 & -15.12 & 1 & 4 & 0.9 (4.48) & 1.1 & 2.33 & -12.79 & 7.7  (38.3) \\
KK\,65        & 7.72 & -14.52 & 0 & 0 & 0     (0)  & 1.2 & 2.34 & -12.18 &   0   (0) \\
UGC\,3755 & 7.41 & -15.50 & 8 & 1 & 5.05 (5.68) & 1.3 & 2.43 & -13.07 & 47.3 (53.3) \\
\tableline
Dregs         &      &       &       &        &       &            &     &    &    &   --   & -- &  --  &   --  & \\
UGC\,9128 & 2.24 & -12.06 & 0 & 0 & 0     (0)  & 0.5 & 2.85 & -9.21 &   0   (0) \\
\tableline\tableline
\end{tabular}
\tablenotetext{a}{\protect\cite{Tully06}}
\tablenotetext{b}{Information for UGC\,9240, which belongs to this group, was taken from \protect\cite{Sharina05}}
\tablenotetext{c}{From RC3}
\end{minipage}
\end{table*}  

\clearpage

\begin{table}
\begin{minipage}{160mm}
\scriptsize
%\vspace{-0.5cm}
\caption{General properties of the GC candidates in the studied dIrrs. According to the colors of the GC candidates, they are separated with lines on blue, red and likely background contamination sub-populations.\label{GCCs}}
\hspace{-0.5cm}
\begin{tabular}{p{1.9cm}cccccccccc}
\tableline\tableline
ID	&	X~,~Y	&	RA\ \ DEC~(2000)	&	$M_V^0$	&    $(V-I)_0$    &    $r_{\rm h}$    &    $r_{\rm h}$    &    $\epsilon$    &    d$_{{\rm proj}}$    &    d$_{{\rm proj}}^{{\rm norm}}$ \\
    &    (pix)    &   (hh:mm:ss)   &    (mag)    &    (mag)    &    (pix)    &    (pc)    &        &    (kpc)    &    (kpc) \\
\tableline
E154-023-01	&	3472.05~,~2329.00 &	02:56:49.17 -54:33:16.51	&	$-4.78\pm0.04$  &	$1.05\pm0.01$	&	$2.02\pm0.09$	&	$4.68\pm0.15$  &       0.10    &       1.99    &       0.75 \\
IC1959-02	&	3083.73~,~1739.38 &	03:33:14.24 -50:25:40.16	&	$-6.19\pm0.05$  &	$1.03\pm0.01$	&	$4.23\pm0.01$	&	$6.21\pm0.11$  &       0.29    &       1.71    &       1.42 \\
IC1959-04	&	2224.47~,~2273.96 &	03:33:12.43 -50:24:52.60	&	$-9.83\pm0.05$  &	$0.97\pm0.01$	&	$2.11\pm0.00$	&	$3.27\pm0.10$  &       0.08    &       0.23    &       0.19 \\
IC1959-05	&	1982.16~,~2506.51 &	03:33:11.51 -50:24:38.34	&	$-6.54\pm0.05$  &	$1.10\pm0.01$	&	$1.81\pm0.00$	&	$2.81\pm0.10$  &       0.15    &       0.26    &       0.22 \\
IC1959-06	&	544.56~,~3554.56 &	03:33:07.70 -50:23:17.16	&	$-7.60\pm0.05$  &	$0.90\pm0.01$	&	$1.72\pm0.02$	&	$2.67\pm0.12$  &       0.11    &       2.87    &       2.39 \\
IC1959-07	&	2852.76~,~1501.31 &	03:33:15.71 -50:25:31.32	&	$-6.11\pm0.06$  &	$0.78\pm0.03$	&	$2.66\pm0.05$	&	$4.12\pm0.15$  &       0.06    &       1.69    &       1.41 \\
IC1959-08	&	3366.00~,~2605.55 &	03:33:09.50 -50:25:45.00	&	$-4.65\pm0.05$  &	$0.99\pm0.01$	&	$1.40\pm0.07$	&	$2.17\pm0.17$  &       0.04    &       1.87    &       1.56 \\
IC1959-09	&	2884.04~,~3886.73 &	03:33:03.47 -50:25:08.15	&	$-5.06\pm0.06$  &	$1.08\pm0.02$	&	$1.67\pm0.05$	&	$2.59\pm0.15$  &       0.32    &       2.48    &       2.07 \\
KK16-01		&	2158.83~,~2265.25 &	01:55:23.51  27:57:29.12	&	$-5.27\pm0.05$  &	$1.05\pm0.01$	&	$2.40\pm0.12$	&	$3.77\pm0.21$  &       0.18    &       1.20    &       2.50 \\
KK27-03		&	3744.75~,~3274.22 &	03:20:50.39 -66:19:38.16	&	$-4.81\pm0.07$  &	$1.02\pm0.01$	&	$1.96\pm0.05$	&	$3.08\pm0.11$  &       0.09    &       1.57    &       7.05 \\
N1311-01	&	493.31~,~415.26 &	03:20:20.05 -52:10:14.95	&	$-5.03\pm0.04$  &	$1.09\pm0.01$	&	$2.06\pm0.06$	&	$3.24\pm0.11$  &       0.08    &       3.57    &       2.95 \\
N1311-04	&	325.69~,~3554.20 &	03:20:04.00 -52:09:21.26	&	$-6.49\pm0.03$  &	$0.82\pm0.01$	&	$3.63\pm0.07$	&	$3.35\pm0.11$  &       0.18    &       3.05    &       2.52 \\
N1311-05	&	2219.00~,~2235.00 &	03:20:07.86 -52:11:11.03	&	$-8.59\pm0.03$  &	$0.91\pm0.01$	&	$2.98\pm0.01$	&	$2.75\pm0.06$  &       0.06    &       0.27    &       0.22 \\
N1311-06	&	1997.27~,~2526.72 &	03:20:06.69 -52:10:56.18	&	$-7.33\pm0.04$  &	$0.80\pm0.01$	&	$8.47\pm0.06$	&	$7.81\pm0.11$  &       0.04    &       0.47    &       0.39 \\
N784-01		&	358.61~,~1688.27 &	02:01:18.09  28:48:42.92	&	$-7.53\pm0.05$  &	$0.97\pm0.01$	&	$2.54\pm0.00$	&	$2.34\pm0.07$  &       0.17    &       2.35    &       0.98 \\
N784-03		&	1788.73~,~2006.70 &	02:01:16.42  28:49:52.81	&	$-6.63\pm0.06$  &	$0.88\pm0.01$	&	$3.07\pm0.00$	&	$2.83\pm0.07$  &       0.12    &       0.59    &       0.25 \\
N784-04		&	3598.94~,~3096.28 &	02:01:16.57  28:51:38.43	&	$-6.38\pm0.07$  &	$0.73\pm0.03$	&	$10.59\pm0.05$  &	$9.76\pm0.12$  &       0.10    &       2.11    &       0.88 \\
N784-09		&	3802.58~,~3811.71 &	02:01:18.54  28:52:05.14	&	$-6.61\pm0.06$  &	$0.77\pm0.02$	&	$3.75\pm0.03$	&	$2.81\pm0.10$  &       0.09    &       2.82    &       1.18 \\
U1281-02	&	2147.49~,~2346.95 &	01:49:32.54  32:35:25.26	&	$-7.58\pm0.04$  &	$0.97\pm0.01$	&	$1.82\pm0.02$	&	$1.36\pm0.06$  &       0.11    &       0.17    &       0.12 \\
U3755-01	&	1638.62~,~2239.86 &	07:13:51.95  10:31:42.11	&	$-9.18\pm0.06$  &	$0.80\pm0.01$	&	$1.55\pm0.01$	&	$1.16\pm0.18$  &       0.20    &       1.30    &       0.96 \\
U3755-02	&	2864.94~,~2681.61 &	07:13:50.61  10:30:40.01	&	$-7.26\pm0.06$  &	$1.01\pm0.01$	&	$1.73\pm0.02$	&	$1.30\pm0.20$  &       0.12    &       1.16    &       0.85 \\
U3755-03	&	1497.29~,~2691.75 &	07:13:50.40  10:31:48.33	&	$-6.58\pm0.06$  &	$0.95\pm0.01$	&	$2.53\pm0.07$	&	$1.90\pm0.25$  &       0.19    &       1.72    &       1.26 \\
U3755-04	&	960.61~,~2756.50 &	07:13:50.11  10:32:15.02	&	$-8.73\pm0.06$  &	$1.02\pm0.01$	&	$1.56\pm0.00$	&	$1.17\pm0.18$  &       0.13    &       2.65    &       1.95 \\
U3755-05	&	2836.61~,~2038.49 &	07:13:52.78  10:30:42.62	&	$-6.64\pm0.08$  &	$1.06\pm0.04$	&	$4.14\pm0.15$	&	$3.10\pm0.33$  &       0.06    &       0.96    &       0.71 \\
U3755-06	&	2003.00~,~2200.00 &	07:13:52.13  10:31:23.97	&	$-8.23\pm0.06$  &	$0.89\pm0.01$	&	$3.11\pm0.00$	&	$2.42\pm0.18$  &       0.04    &       0.66    &       0.48 \\
U3755-07	&	2236.00~,~2229.53 &	07:13:52.06  10:31:12.28	&	$-7.33\pm0.07$  &	$0.86\pm0.01$	&	$3.02\pm0.00$	&	$2.35\pm0.18$  &       0.08    &       0.24    &       0.18 \\
U3755-08	&	2529.77~,~2427.00 &	07:13:51.43  10:30:57.23	&	$-8.25\pm0.07$  &	$0.91\pm0.01$	&	$4.53\pm0.00$	&	$5.98\pm0.18$  &       0.19    &       0.40    &       0.30 \\
U3974-01	&	964.54~,~1568.30 &	07:41:59.49  16:49:00.92	&	$-7.20\pm0.04$  &	$0.85\pm0.01$	&	$4.93\pm0.05$	&	$5.41\pm0.19$  &       0.16    &       3.06    &       0.87 \\
U3974-02	&	1910.55~,~1800.60 &	07:41:58.16  16:48:16.12	&	$-8.76\pm0.04$  &	$0.97\pm0.01$	&	$1.65\pm0.01$	&	$1.50\pm0.15$  &       0.14    &       1.57    &       0.44 \\
U3974-03	&	2419.29~,~1739.74 &	07:41:58.08  16:47:50.52	&	$-7.86\pm0.04$  &	$0.94\pm0.01$	&	$2.13\pm0.02$	&	$1.94\pm0.16$  &       0.13    &       1.67    &       0.47 \\
U3974-04	&	2485.60~,~2663.96 &	07:41:54.87  16:47:54.72	&	$-8.27\pm0.05$  &	$0.99\pm0.03$	&	$1.96\pm0.01$	&	$1.78\pm0.15$  &       0.11    &       0.63    &       0.18 \\
U4115-01	&	1217.34~,~3760.26 &	07:57:03.79  14:22:41.01	&	$-7.61\pm0.05$  &	$0.96\pm0.01$	&	$2.16\pm0.02$	&	$1.96\pm0.18$  &       0.06    &       1.39    &       0.91 \\
U685-01		&	2866.04~,~2345.56 &	01:07:26.18  16:40:56.84	&	$-7.01\pm0.03$  &	$0.94\pm0.01$	&	$3.03\pm0.02$	&	$5.75\pm0.06$  &       0.02    &       1.31    &       1.88 \\
U685-03		&	3015.40~,~2596.32 &	01:07:25.68  16:40:44.19	&	$-7.94\pm0.04$  &	$0.98\pm0.02$	&	$4.34\pm0.08$	&	$5.70\pm0.11$  &       0.07    &       1.22    &       1.75 \\
U685-04		&	2066.37~,~2746.89 &	01:07:23.60  16:41:21.88	&	$-8.63\pm0.03$  &	$0.91\pm0.01$	&	$2.13\pm0.00$	&	$2.80\pm0.04$  &       0.13    &       0.61    &       0.88 \\
U685-05		&	2993.85~,~2926.33 &	01:07:24.64  16:40:37.05	&	$-7.82\pm0.03$  &	$0.97\pm0.01$	&	$3.75\pm0.02$	&	$3.59\pm0.06$  &       0.04    &       1.01    &       1.44 \\
U685-06		&	1991.53~,~3155.90 &	01:07:22.24  16:41:15.14	&	$-8.35\pm0.03$  &	$0.90\pm0.01$	&	$2.01\pm0.00$	&	$1.78\pm0.04$  &       0.06    &       0.25    &       0.36 \\
\tableline
E154-023-02	&	500.587~,~3732.66	&	02:57:01.03 -54:35:24.44	&	$-4.78\pm0.03$  &	$1.13\pm0.01$	&	$1.87\pm0.02$	&	$3.85\pm0.03$	&	0.15	&	2.89	&	1.09 \\
E154-023-03	&	750.00~,~3799.00	&	02:57:01.06 -54:35:11.54	&	$-4.39\pm0.03$  &	$1.28\pm0.01$	&	$2.59\pm0.05$	&	$5.32\pm0.07$	&	0.09	&	2.71	&	1.02 \\
KK17-01		&	1821.97~,~1410.22	&	02:00:15.85  28:50:42.59	&	$-4.58\pm0.08$  &	$1.22\pm0.02$	&	$2.02\pm0.03$	&	$3.56\pm0.03$	&	0.03	&	2.12	&	8.41 \\
KK27-02		&	2665.38~,~3563.55	&	03:20:57.69 -66:19:03.74	&	$-6.15\pm0.08$  &	$1.20\pm0.02$	&	$4.55\pm0.01$	&	$6.8\pm0.01$	&	0.11	&	0.58	&	2.61 \\
N1311-02	&	1570.35~,~490.91	&	03:20:17.96 -52:11:05.37	&	$-5.69\pm0.03$  &	$1.39\pm0.01$	&	$2.13\pm0.02$	&	$4.17\pm0.02$	&	0.11	&	2.72	&	2.24 \\
N784-07		&	1398.57~,~2354.43	&	02:01:18.31  28:49:44.61	&	$-5.71\pm0.06$  &	$1.22\pm0.03$	&	$3.01\pm0.08$	&	$5.62\pm0.10$	&	0.09	&	0.88	&	0.37 \\
N784-08		&	1020.28~,~3049.7	&	02:01:21.32  28:49:45.61	&	$-4.38\pm0.07$  &	$1.14\pm0.04$	&	$2.79\pm0.08$	&	$5.2\pm0.10$	&	0.16	&	1.61	&	0.67 \\
U3755-09	&	2247.0~,~2767.67	&	07:13:50.24  10:31:10.73	&	$-6.62\pm0.06$  &	$1.22\pm0.01$	&	$5.05\pm0.22$	&	$13.44\pm0.40$  &	0.12	&	0.90	&	0.66 \\
U4115-02	&	2306.86~,~890.00	&	07:56:54.05  14:23:40.30	&	$-5.72\pm0.05$  &	$1.20\pm0.01$	&	$3.09\pm0.03$	&	$8.55\pm0.06$	&	0.16	&	4.71	&	3.07 \\
U4115-03	&	214.30~,~2458.00	&	07:56:59.20  14:21:53.09	&	$-6.44\pm0.05$  &	$1.23\pm0.01$	&	$3.57\pm0.01$	&	$9.88\pm0.01$	&	0.14	&	3.49	&	2.28 \\
U4115-04	&	3215.00~,~2926.0	&	07:57:01.16  14:24:22.25	&	$-4.94\pm0.07$  &	$1.26\pm0.04$	&	$2.17\pm0.01$	&	$6.01\pm0.02$	&	0.28	&	2.70	&	1.76 \\
U4115-05	&	925.00~,~3822.00	&	07:57:03.97  14:22:26.30	&	$-4.86\pm0.06$  &	$1.35\pm0.02$	&	$1.54\pm0.05$	&	$4.26\pm0.10$	&	0.18	&	1.92	&	1.25 \\
U8760-01	&	2932.65~,~3042.88	&	13:50:50.73  38:01:48.27	&	$-4.80\pm0.07$	&	$1.28\pm0.02$	&	$3.60\pm0.07$	&	$4.16\pm0.05$	&	0.22	&	0.51	&	0.89 \\
\tableline
E115-021-01	&	505.00~,~1425.43	&	02:37:54.24 -61:18:45.39        &       $-4.78\pm0.03$  &       $1.48\pm0.01$   &       $2.74\pm0.03$   &       $4.91\pm0.04$   &       0.12    &       2.10    &       1.52 \\
IC1959-01	&	395.93~,~1110.70	&	03:33:20.36 -50:23:35.17        &       $-5.43\pm0.05$  &       $1.46\pm0.01$   &       $3.16\pm0.03$   &       $6.88\pm0.04$   &       0.06    &       3.13    &       2.61 \\
IC1959-03	&	670.10~,~1963.78	&	03:33:15.70 -50:23:39.77        &       $-5.42\pm0.05$  &       $1.54\pm0.01$   &       $2.23\pm0.02$   &       $4.85\pm0.04$   &       0.18    &       2.20    &       1.83 \\
KK27-01		&	2377.27~,~3224.14	&	03:21:01.00 -66:19:13.72	&	$-5.22\pm0.08$	&	$1.59\pm0.01$	&	$2.84\pm0.08$	&	$4.25\pm0.08$	&	0.04	&	0.20	&	0.88 \\
KK65-01		&	2016.4~,~2749.84	&	07:42:39.75  16:34:37.65	&	$-6.12\pm0.08$	&	$1.48\pm0.02$	&	$2.75\pm0.01$	&	$4.32\pm0.01$	&	0.21	&	0.88	&	0.87 \\
N1311-03	&	2869.85~,~2768.65	&	03:20:04.05 -52:11:34.39	&	$-6.47\pm0.03$  &	$1.66\pm0.01$	&	$2.63\pm0.01$	&	$5.14\pm0.01$	&	0.16	&	0.86	&	0.71 \\
N784-02		&	3337.92~,~1645.64	&	02:01:12.29  28:50:50.86	&	$-6.09\pm0.05$  &	$1.57\pm0.01$	&	$1.83\pm0.01$	&	$3.41\pm0.01$	&	0.13	&	1.80	&	0.75 \\
U685-02		&	3403.67~,~2455.21	&	01:07:26.77  16:40:30.71	&	$-4.66\pm0.03$  &	$1.74\pm0.01$	&	$1.91\pm0.03$	&	$3.22\pm0.04$	&	0.06	&	1.68	&	2.40 \\
U1281-01	&	1439.38~,~3836.35	&	01:49:38.95  32:35:09.73	&	$-7.40\pm0.03$  &	$1.25\pm0.01$	&	$2.75\pm0.02$	&	$5.05\pm0.03$	&	0.10	&	2.13	&	1.53 \\
U3974-05	&	1729.83~,~2874.48	&	07:41:54.57  16:48:33.71	&	$-8.71\pm0.05$  &	$1.33\pm0.02$	&	$3.08\pm0.01$	&	$8.88\pm0.01$	&	0.11	&	1.07	&	0.30 \\
\tableline
\end{tabular}
\end{minipage}
\end{table}

\begin{table}
\caption{Results of K-S tests for $V-I$, $M_V$, $r_{\rm h}$ and $\epsilon$ between our bGCs and the old GCs in the LMC and our Galaxy.}\label{k-s-results}
\begin{tabular}{c|cccc|cccc}
\tableline\tableline
& \multicolumn{4}{|c|}{bGCs} & \multicolumn{4}{|c|}{LMC} \\
& $V-I$ & $M_V$ & $r_{\rm h}$ & $\epsilon$ & $V-I$ & $M_V$ & $r_{\rm h}$ & $\epsilon$ \\
\tableline
YH & 54\% & 50\% & 1\% & 1\% & 25\% & 34\% & 32\% & 3\% \\
OH & 1\% & 19\% & 86\% & 3\% & 1\% & 78\% & 7\% & 20\% \\
LMC & 73\% & 66\% & 8\% & 99\% & -- & -- & -- & -- \\
\tableline
\end{tabular}
\end{table}

\end{document}